\documentclass[12pt,letter]{article}
\usepackage[left=.5in,right=.5in,top=1.5in,bottom=1.5in]{geometry}
\usepackage{amsmath}
\usepackage{amsthm}
\usepackage{mathpazo}
\usepackage{setspace}
\usepackage{graphicx}
\usepackage{xcolor}
\usepackage{times}
\usepackage{placeins}
\usepackage{subcaption, booktabs}
\usepackage{hyperref}
\usepackage{authblk} 
\usepackage{array}
\usepackage{amsthm}
\usepackage{enumerate}
\usepackage{comment}
\usepackage{lineno}
\usepackage{blkarray}
\newcolumntype{H}{>{\setbox0=\hbox\bgroup}c<{\egroup}@{}}
\graphicspath{{./plots/}}
\usepackage{float}
\usepackage{bm}
\usepackage{natbib}
\usepackage{natbib}
\newcommand{\norm}[1]{\lVert#1\rVert}
\DeclareMathOperator{\Var}{\text{Var}}
\DeclareMathOperator{\Cov}{\text{cov}}
\DeclareMathOperator*{\argmin}{arg\,min}
\newcommand{\supp}{\text{supp}}
\newcommand{\bXhat}{\widehat{\mathbf{X}}}
\newcommand{\bv}{\mathbf{v}}
 
\newcommand{\bU}{\mathbf{U}}
\newcommand{\bA}{\mathbf{A}}
\newcommand{\bD}{\mathbf{D}}

\newcommand{\bZ}{\mathbf{Z}}
\newcommand{\bz}{\mathbf{z}}
\newcommand{\bI}{\mathbf{I}}
\newcommand{\bE}{\mathbf{E}}
\newcommand{\bGamma}{\bm{\Gamma}}
\newcommand{\bgamma}{\bm{\gamma}}
\newcommand{\bbeta}{\bm{\beta}}
\newcommand{\bB}{\mathbf{B}}
\newcommand{\bW}{\mathbf{W}}
\newcommand{\bX}{\mathbf{X}}
\newcommand{\bx}{\mathbf{x}}

\newcommand{\by}{\mathbf{y}}
\newcommand{\bby}{{y}}
\newcommand{\boldv}{\textbf{v}}

\newcommand*{\bC}{\mathbf{C}}
\newcommand*{\bQ}{\mathbf{Q}}

\newcommand*{\bM}{\mathbf{M}}

\newcommand*{\bP}{\mathbf{P}}
\newcommand*{\bH}{\mathbf{H}}
\newcommand*{\bJ}{\mathbf{J}}

\newcommand{\boldeta}{\bm\eta}
\newcommand{\bdelta}{\bm\delta}
\newcommand{\bSigma}{\bm\Sigma}
\newcommand{\bLambda}{\bm\Lambda}
\newcommand{\mcX}{\mathcal{X}}
\newcommand{\mcZ}{\mathcal{Z}}
\newcommand{\mcV}{\mathcal{V}}
\newcommand{\mcT}{\mathcal{T}}
\newcommand{\mcW}{\mathcal{W}}
\newcommand{\mcU}{\mathcal{U}}
\newcommand{\tr}{\text{trace}}

\newcommand{\bOmega}{\bm\Omega}
\newcommand{\bT}{T}
\usepackage{xspace}
\newcommand{\SIR}{SIR\xspace}
\newcommand{\SDR}{SDR\xspace}
\newcommand{\T}{\top}
\newtheorem{condition}{Condition}
\newcommand{\RE}{restricted eigenvalue\xspace}
\newcommand*{\bXi}{\bm{\Xi}}
\newcommand*{\be}{\bm{e}}
\newcommand{\bhatLambda}{\bm{\widehat\Lambda}}
\newcommand*{\spn}{\text{span}}

\newtheorem{proposition}{Proposition}
\newtheorem{lemma}{Lemma}
\newtheorem{theorem}{Theorem}

\makeatletter
\g@addto@macro\normalsize{%
  \setlength\abovedisplayskip{8pt}
  \setlength\belowdisplayskip{8pt}
  \setlength\abovedisplayshortskip{8pt}
  \setlength\belowdisplayshortskip{8pt}
}
\makeatother

\newcommand{\blind}{0}
\begin{document}

\if0\blind
{
  \title{High-dimensional sliced inverse regression with endogeneity} 
\author[1]{Linh H. Nghiem\thanks{Corresponding author: linh.nghiem@sydney.edu.au}} 
\author[2]{Francis K.C. Hui} 
\author[3]{Samuel Muller} 
\author[4]{A.H. Welsh} 
\affil[1]{School of Mathematics and Statistics, University of Sydney, Australia}
\affil[2]{Research School of Finance, Actuarial Studies and Statistics, The Australian National University, Australia}
\affil[3]{School of Mathematical and Physical Sciences, Macquarie University, Sydney Australia}
\date{}
  \maketitle
} \fi

\if1\blind
{
  \bigskip
  \bigskip
  \bigskip
  \begin{center}
    {\LARGE\bf High-dimensional sliced inverse regression with endogeneity}
\end{center}
  \medskip
} \fi

\bigskip

\thispagestyle{empty} 

\begin{abstract}
Sliced inverse regression is a popular sufficient dimension reduction method that identifies a small number of linear transformations of the covariates without losing regression information with the response. In high-dimensional settings, it can be combined with sparsity penalties to achieve simultaneous sufficient dimension reduction and variable selection. 
Both classical and sparse \SIR estimators assume the covariates are exogenous. However, endogeneity can arise in a variety of situations, such as when variables are omitted or are measured with error.
In this article, we show that such situations 
invalidate \SIR and lead to inconsistent estimation of the true central subspace. To address this challenge, we propose a two-stage lasso \SIR estimator, which first constructs a sparse high-dimensional instrumental variables model to obtain fitted values of the covariates spanned by the instruments, and then applies \SIR augmented with a lasso penalty to these fitted values. We establish theoretical bounds for the estimation and selection consistency of the true dimension reduction subspace for the proposed two-stage lasso \SIR estimator, allowing the number of covariates and instruments to grow exponentially with the sample size. Simulation studies and applications to two real-world datasets in nutrition and genetics illustrate the superior empirical performance of the proposed estimator, compared with existing methods which ignore endogeneity and/or nonlinearity in the outcome model.


\end{abstract}
\newpage

\section{Introduction} \label{sec:intro}
Sufficient dimension reduction \citep[SDR]{ma2013review,li2018sufficient} refers to a class of methods that assume the outcome $y \in \mathbb{R}$ depends on a set of random covariates $\bX \in \mathbb{R}^{p}$ only through a small number of linear combinations $\bbeta_1^\T \bX, \ldots, \bbeta_d^\T \bX$, with $d\ll p$. These linear combinations are known as sufficient predictors, and retain the full regression information between the response and covariates. More specifically, letting $\bB = [\bbeta_1, \ldots, \bbeta_d]$ denote a $p \times d$ matrix, we can formulate SDR as a multiple index model 
\begin{equation}
y = f(\bB^\T \bX, \varepsilon),
\label{eq:multipleindex}
\end{equation} 
where $f$ denotes an unknown link function, and $\varepsilon \in \mathbb{R}$ is random noise. The target of estimation is then the column space of $\bB$, also known as the central subspace. 
Since the pioneering work of \citet{li1991sliced}, a vast literature has developed on different sufficient dimension reduction approaches, ranging from inverse-moment-based and regression-based methods \citep[e.g.,][]{cook2008, bura2022sufficient}, to forward regression \citep[e.g.,][]{xia2002adaptive,dai2024new} and semiparametric techniques \citep[e.g.,][]{zhou2016principal, sheng2016sufficient}. Among the various SDR methods, \SIR \citep{li1991sliced} remains the most widely used due to its simplicity and computational efficiency, and as such continues to be the subject of much research \citep[see][among many others]{tan2018convex, xu2022distributed}. When the number of covariates $p$ is large, it is often (further) assumed each linear combination depends only on a few covariates, i.e., the vectors $\bbeta_1, \ldots, \bbeta_d$ are sparse. Recent literature has thus sought to combine \SIR (and other \SDR methods) with variable selection to achieve sparse sufficient dimension reduction \citep[e.g.,][]{lin2018consistency, lin2019sparse,nghiem2021sparse}.

Much of the literature dedicated to \SIR and many other \SDR methods, including those mentioned above, has focused on the setting where all covariates are exogenous, i.e. in model \eqref{eq:multipleindex}, $\bX$ and $\varepsilon$ are independent,
implying $y \perp \bX \mid \bB^\T \bX$, where `$\perp$' denotes statistical independence. However, this assumption may not be reasonable in settings where the random noise $\varepsilon$ is dependent on $\bX$, as in the case where the covariates are endogenous.  Little has been done to tackle sufficient dimension reduction with endogeneity, with the following two common scenarios. %
First, \textit{omitted-variable bias} occurs when one or more (truly important) covariates is not included in the model. In Section \ref{subsection:data2} for instance, we consider a genomic study relating a biological outcome (e.g., body weight) to gene expressions of mice. Outside of gene expressions, there can be potentially unmeasured confounding factors, such as those related to environmental conditions, which influence both gene expression levels and the outcome.    
Second, \textit{covariate measurement error} arises when one or more of the covariates are not measured accurately. For instance, in a second data application in Section \ref{subsec:data1}, we consider a nutritional study relating an individual's health outcomes to their long--term diet, collected through 24--hour recall interviews that are well-known to be subject to measurement errors \citep{carroll2006measurement}. Let $\bX$ denote the long-term dietary covariates. The observed data can then be modeled as $\bW = \bX + \bE$, with $\bE$ denoting an additive measurement error independent of $\bX$. In this situation, even when in the true model \eqref{eq:multipleindex} it holds that  $\varepsilon \perp \bX$, a naive analysis which replaces $\bX$ by $\bW$ results in endogeneity, since $y = f(\bB^\T \bW+\bB^\T \bE, \varepsilon)$ and $\bE$ is correlated with $\bW$. This causes naive estimators that ignore measurement errors to be inconsistent. 

In this article, we show under general conditions that endogeneity invalidates \SIR, leading to inconsistent estimates of the true column space of $\bB$. To address this challenge, we propose a two-stage lasso \SIR method to estimate the central space in the presence of endogeneity, when valid instrumental variables are available. In the first stage, we construct a sparse multivariate linear model regressing all the covariates against the instruments, and  obtain fitted values from this model. In the second stage, we apply a sparse \SIR between the response and the fitted values from the first stage to perform simultaneous SDR and variable selection. 

Addressing endogeneity via instrumental variables has been extensively studied in the econometrics and statistics literature; see \citet{angrist2001instrument} for a review. 
However, classical instrumental variable estimators are mostly studied with a linear model for the outcome, i.e., $y = \bX^\T \bbeta_0 + \varepsilon$, and usually compute a two-stage least squares or a limited information maximum likelihood estimator. \citet{hansen2008estimation} showed in this setting that these estimators are consistent for $\bbeta_0$ when the number of instruments $q$ increases slowly with the sample size. More recently, a number of methods have combined instrumental variables with regularization techniques on the covariates and/or instrumental variables to improve their performance in high-dimensional settings e.g., the lasso-type generalized methods of moments \citep{caner2009lasso, caner2018adaptive}, focused generalized methods of moments \citep{fan2014endogeneity}, and two-stage penalized least squares \citep{lin2015regularization}; the last of these methods inspires this article. Another line of research has focused on the use of sparsity-inducing penalties to estimate optimal instruments from a large set of a-priori valid instruments or some variation thereof \citep[e.g.,][]{belloni2012sparse, chernozhukov2015post,windmeijer2019use}.
To the best of our knowledge though, none of these methods allow for the presence of non-linear link functions in the model for the response e.g., the multiple index model in \eqref{eq:multipleindex}. While there is some related literature on single index models with endogeneity \citep{hu2015identification, gao2020endogeneity}, their approach requires approximation of the link function and is not easy to generalize to multiple index models.

Under a setting where both the number of covariates $p$ and the number of instrumental variables $q$ can grow exponentially with sample size $n$, we derive theoretical bounds on the estimation and selection consistency of our proposed two-stage lasso \SIR estimator. Simulation studies demonstrate the proposed estimator exhibits superior finite sample performance compared with a single-stage lasso \SIR estimator when endogeneity is present but ignored, and the two-stage regularized estimator of \citet{lin2015regularization} which accounts for endogeneity but assumes the outcome model is linear in nature. We apply the proposed method to two aforementioned real data applications studying the effects of different nutrients on the cholesterol level (covariate measurement error), and for identifying genes that are relevant to the body weight of mice (omitted variable bias). 

We establish some notation to be used throughout the paper. For any matrix $\bA$ with elements $a_{ij}$, let $\Vert \bA \Vert_1 = \max_j \sum_{i}\vert  a_{ij} \vert$, $\Vert \bA \Vert_\infty = \max_i \sum_{j} \vert a_{ij} \vert$, and $\norm{\bA}_{\max} = \max_{i,j} \vert a_{i,j} \vert$ denote the matrix $1$-norm, $\infty$-norm, and element-wise maximum norm, respectively. Also, let $\Vert \bA \Vert_2$ denote the $\ell_2$-operator norm, i.e. the largest singular value of $\bA$. For a square matrix $\bA$, we use $\lambda_j(\bA)$ to denote the $j$th largest eigenvalue of $\bA$, whose corresponding eigenvector is given by $\boldeta_j(\bA)$. When $j=1$, we suppress the subscript and write $\lambda(\bA)$ and $\boldeta(\bA)$. For any vector $\bv$ with elements $v_i$, let $\Vert \bv \Vert_2 = \left(\sum_{i} v_i^2 \right)^{1/2}$, $\Vert \bv\Vert_1 = \sum_{i} \vert v_i \vert$, and $\Vert \bv \Vert_\infty = \max_{i} \vert v_i \vert $ denote its Euclidean norm, $\ell_1$ norm, and $\ell_\infty$ norm, respectively. Finally, for an index set $I$, let $\bv_I$ denote the subvector of $\bv$ formed with the elements of $\bv$ whose indices $i\in I$, $\vert I \vert$ denote its cardinality, and $I^c$ denote its complement. 
\section{\SIR and endogeneity}
\label{section:review}
In this section, we first establish the setting of interest, namely a high-dimensional multiple index model for the response along with a high-dimensional linear instrumental variable model for the covariates characterizing endogeneity. We then review \SIR and a version augmented with the lasso penalty (Section \ref{subsection:review}), and then demonstrate inconsistency of \SIR under endogeneity in Section \ref{subsec:inconsistency}.

\subsection{Review of \SIR and lasso \SIR}
\label{subsection:review}

Consider a set of $n$ independent observations $\{(y_i, \bX_i^\T, \bZ_i^\T); i = 1,\ldots,n\}$ where, 
without loss of generality, we assume $E(\bX_i) = E(\bZ_i) = {0}$. Each triplet $(y_i, \bX_i^\T, \bZ_i^\T)$ follows the multiple index model \eqref{eq:multipleindex} coupled with a linear instrumental variable model. That is, the full model is given by
\begin{align}
y_i = f(\bX_i^\T \bB, ~\varepsilon_i), ~\quad  \bX_i = \bGamma^\T\bZ_i + \bU_i, ~\quad i=1,\ldots ,n,
\label{eq:multipleindex_linearinstrument}
\end{align}
where $\bB \in \mathbb{R}^{p\times d}$, $\bGamma \in \mathbb{R}^{q \times p}$,  $\varepsilon_i \in \mathbb{R}$ denotes the random noise for the outcome model, and $\bU_i \in \mathbb{R}^{p}$ denotes the random noise for the instrumental variable model. The vector $(\bU_i^\T, \varepsilon_i)^\T \in \mathbb{R}^{p+1}$ is assumed to have mean zero and a finite covariance matrix $\bSigma$ whose diagonal elements are denoted by $\sigma_j^2$ for $j=1,\ldots, p+1$. To capture endogeneity between $\bX$ and $\varepsilon$, the off-diagonal elements in the last row and column of $\bSigma$ are generally non-zero. The instrumental variable $\bZ_i$ satisfies $\bZ_i \perp (\bU_i^\T, \varepsilon_i)^\T$ for all $i = 1,\ldots,n$. 

We aim to estimate the column space of $\bB$ when the dimensions $p$ and $q$ grow exponentially with the sample size $n$. A common assumption in such settings is that sparsity holds.  Specifically: 
(i) each covariate $X_j$ depends upon only a few instrumental variables, such that in the instrumental variable model the $j$th column of $\bGamma$ is a sparse vector with at most $r_j < q$ non-zero elements, where $r = \max_{j=1,\ldots, p} r_j $ is bounded, and (ii) the multiple index model is sparse, such that $y$ depends only on $s < p$ covariates i.e., at most $s$ rows of $\bB$ are non-zero row vectors. Equivalently, if $\mathcal{P}(\bB) = \bB(\bB^\T\bB)^{-1}\bB^\T$ denotes the corresponding projection matrix, then only $s$ diagonal elements of $\mathcal{P}(\bB)$ are non-zero. Without loss of generality, we assume $S = \{1, \ldots, s\}$ i.e., the first $s$ covariates indexes this set of non-zero elements, or equivalently the support set of the column space of $\bB$.

Let $\mcX$ denote the $n \times p$ matrix of covariates, whose $i$th rows is $\bX_i^\T$. Similarly, let $\mcZ$ denote the $n\times q$ matrix of instrumental variables whose $i$th row is  $\bZ_i^\T$. Also, let $\bx_j$ and $\bz_j$ denote the $j$th column of $\mcX$ and $\mcZ$, respectively.  For the remainder of this article, we use $y, \bX$, and $\bZ$ to denote the random variables from which $(y_i, \bX_i^\T, \bZ_i^\T)$ is drawn.

The \SIR estimator \citep{li1991sliced} of the column space of $\bB$ was developed under the assumption that the random noise $\varepsilon$ is independent of the covariates $\bX$. 
Furthermore, a key prerequisite of \SIR is the following linearity condition, which is satisfied when the covariates $\bX$ follow an elliptical distribution.

\begin{condition}
The conditional expectation $E(\bX\mid \bB^\T \bX)$ is a linear function of $\bB^\T \bX$.
\label{eq:linearcondition}
\end{condition}

\noindent Under this condition, the population foundation for \SIR is that the inverse regression curve $E(\bX \mid y)$ is contained in the linear subspace spanned by $\bSigma_{\bX} \bB$ \citep[Theorem 3.1,][]{li1991sliced}. Consequently, if we let $\{\boldeta^*_1, \ldots, \boldeta^*_d\} = \{\boldeta_1(\bLambda^*), \ldots, \boldeta_d(\bLambda^*)\}$ denote the set of eigenvectors associated with the $d$ largest eigenvalues of $\Lambda^* = \Cov\left\{E(\bX \mid y) \right\}$, then $\bSigma_{\bX} \bbeta_k \propto \boldeta^*_k$ for $k=1,\ldots, d$. Since $\bLambda^* = \bSigma_{\bX} - E\left\{\Cov(\bX\mid y) \right\}$, we can estimate $\bLambda^*$ from some estimates of $\bSigma_{\bX}$ and $\bT=E\left\{\Cov(\bX\mid y) \right\}$: this approach to \SIR is recommended by \citet{li1991sliced}, and used by many authors in the literature \citep{tan2018convex, zou06}, although we acknowledge it is not the only approach possible \citep{lin2019sparse}.

In more detail, suppose we estimate $\bSigma_{\bX}$ by the sample covariance matrix $\widehat{\bSigma}_{\bX} = n^{-1} \mcX^\T \mcX$. When $n <p$, this sample covariance matrix is not invertible, but critically the lasso \SIR estimator defined below does not require such invertibility, as we will see shortly. Next, we partition the data into $H$ non-overlapping slices $S_1, \ldots, S_H$ based on the order statistics of $y$, and compute the sample covariance matrices within each slice and average them as follows
\begin{align*}
\widehat{{T}} = \dfrac{1}{H} \sum_{h=1}^{H} \left\{ \frac{1}{c-1} \sum_{i \in S_h} (\bX_{i} - \bar{\bX}_h) (\bX_{i} - \bar{\bX}_h)^\T\right\}, \text{where} \; \bar{\bX}_h = \dfrac{1}{c} \sum_{i \in S_h} \bX_i,    
\end{align*}
where $c$ denotes the slice size and we have assumed $n = cH$ to simplify the notation. \citet{zhu2006sliced} showed that $\widehat{{T}}$ can be rewritten in the form 
\begin{align*}
\widehat{{T}} = \dfrac{1}{H(c-1)} \mcX^\T (\bI_H \otimes \bP_c) \mcX, \text{where} \; \bP_c = \bI_c - \dfrac{1}{c} {\be}_c{\be}_c^\T,
\end{align*}
and ${e}_c$ denotes a $c\times 1$ vector of ones. We can then estimate $\bLambda^*$ by $\widehat{\bLambda}^* = \widehat{\bSigma}_{\bX} - \widehat{{T}} = n^{-1}\mcX^\T \bD \mcX$, where $\bD = \bI_n - \{c/(c-1)\} (\bI_H \otimes \bP_c)$. 
It follows that the eigenvector $\boldeta^*_k$ can be estimated by $\hat{\boldeta}^*_k = \boldeta_k(\widehat{\bLambda}^*)$, that is, the eigenvector associated with the $k$th largest eigenvalue $\hat\lambda^*_k = \lambda_k(\widehat{\bLambda}^*)$ of $\widehat{\bLambda}^*$ for $k=1,\ldots, d$.

For settings where $p$ is large and $\bbeta_k$ is sparse, consider replacing the population relationship $\bSigma_{\bX}\bbeta_k \propto \boldeta^*_k $ by its sample counterpart. Then we have $\mcX^\T \mcX \bbeta_k \propto \hat{\boldeta}^*_k = (\hat{\lambda}^*_k)^{-1}\mcX^\T\bD\mcX\hat{\boldeta}^*_k$, from which \citet{lin2019sparse} proposed the lasso \SIR estimator as
\begin{equation}
\hat{\bbeta}_k^* =  \arg\!\min_{\bbeta} \frac{1}{2n} \norm{\tilde{{\by}}^*_k - \mcX \bbeta}_2^2 + \mu_k \|{\beta}\|_1,
\label{eq:SIR}
\end{equation}
where $\tilde{{\by}}^*_k = (\hat\lambda^*_k)^{-1} \bD \mcX \hat{\boldeta}^*_k$ is a pseudo-response vector and $\mu_k > 0$ denotes a tuning parameter  for the $k$th pseudo-response. Note the lasso penalty can be replaced by a more general sparsity-inducing penalty \citep[e.g., the adaptive lasso penalty in][]{nghiem2021sparse}, but such generalizations are outside the scope of this article. More importantly, with an appropriate choice of $\mu_k$ the solution $\hat{\bbeta}_k$ is sparse, thus facilitating variable selection for each dimension of the estimator. Finally, variable selection for the estimated dimension reduction subspace is obtained by taking the union of the estimated support sets across all dimensions $k=1,\ldots, d$. 

\subsection{Inconsistency of \SIR under endogeneity} \label{subsec:inconsistency}
When endogeneity is present, the inverse regression curve $E(\bX \mid y)$ is no longer guaranteed to lie in the subspace spanned by $\bSigma_X \bB$, even when Condition \ref{eq:linearcondition} holds for the marginal $\bX$. 
The following proposition establishes this result in the case  $(\bX, \varepsilon)^\T$ follows a multivariate Gaussian distribution.

\begin{proposition}
Assume $(\bX^\T, \varepsilon)^\T$ follows a $(p+1)-$dimensional Gaussian distribution, and let $\bSigma_{\bX, \varepsilon} = \Cov(\bX, \varepsilon)$. Then the inverse regression curve $E(\bX \mid y)$ is contained in the linear subspace spanned by $\bSigma_{\bX} \bB$ if and only if $\bSigma_{\bX}^{-1}\bSigma_{\bX, \varepsilon} \in \text{col}(\bB)$, where $\text{col}(\bB)$ denotes the column space of $\bB$. 
\label{proposition:inconsistency}
\end{proposition}

In Proposition \ref{proposition:inconsistency} the matrix $\bSigma_{\bX}^{-1} \bSigma_{\bX, \varepsilon}$ is the coefficient vector of the best linear predictor of $\varepsilon$ from $\bX$. Hence this proposition implies that, when endogeneity is present, the population foundation for \SIR can \emph{only} hold when the random noise can be predicted from the linear combination $\bB^\T \bX$ itself. This condition is obviously restrictive, as it requires the same set of linear combinations to be used to predict both the response $y$ and the random noise $\varepsilon$.

Unsurprisingly, when the condition in Proposition \ref{proposition:inconsistency} is not satisfied, then the lasso \SIR estimator in \eqref{eq:SIR} is also generally not a consistent estimator of the column space of $\bB$. The following proposition makes this precise in the simple setting where the number of covariates $p$ is fixed, and the penalization is asymptotically negligible.

\begin{proposition} \label{proposition:lassoSIR_inconsistency}
Let $\widehat{\bB}^*$ denote the lasso \SIR estimator formulated in \eqref{eq:SIR}, with tuning parameters $\mu_k \to 0$ for all $k=1,\ldots, d$ as $n  \to \infty$. Then, $\mathcal{P}(\widehat{\bB}^*) \to \mathcal{P}(\bB^*)$, where $\bB^* = \bSigma_X^{-1} \boldeta^*$, where $\boldeta^*$ is a $p\times d$ matrix of eigenvectors of $\Lambda^* = \Cov\left\{E(\bX \mid y)\right\}$. It follows that if the covariance vector $\Sigma_{\bX, \varepsilon} \notin \bSigma_{\bX}\text{col}(\bB)$, then $\Vert P(\widehat{\bB}^*) - P(\bB) \Vert \neq o_p(1)$ as $n \to \infty$ and $\widehat\bB^*$ is inconsistent for the column space of $\bB$.
\end{proposition}

Propositions \ref{proposition:inconsistency} and \ref{proposition:lassoSIR_inconsistency} imply that when the covariance vector $\bSigma_{\bX, \varepsilon} \notin \bSigma_{\bX}\text{col}(\bB)$, then endogeneity invalidates (lasso) \SIR as an estimator of the overall dimension reduction subspace. Note this does not necessarily imply selection inconsistency, i.e. inconsistency in estimating the support set $S = \{1, \ldots, s\}$.
Indeed, selection consistency can still be achieved if the column space of $\bB^*$ has the same support as that of $\bB$; see Appendix A for an empirical demonstration. 
On the other hand, when the number of covariates $p$ diverges with $n$, then it is unlikely $\bB^*$  will share the same support as $\bB$. This motivates us to propose a new \SIR method to handle endogeneity in high-dimensional settings.

\section{A two-stage lasso \SIR estimator}
\label{section:instr_dr}
To overcome the inconsistency of the \SIR estimator when endogeneity is present, we propose a two-stage approach for \SDR and variable selection. Let $\widehat{\bX} = E(\bX \mid \bZ) = \bGamma^\T \bZ$ denote the conditional expectation of the covariates on the instruments. We first impose a new linearity condition to ensure the conditional expectation $E(\widehat{\bX} \mid y)$ behaves similarly to $E(\bX \mid y)$ when no endogeneity is present.

\begin{condition}
The conditional expectation $E(\bXhat \mid \bB^\T \bX)$ is a linear function of $\bB^\T \bX$.
\label{condition:cond2}
\end{condition}

\noindent Despite being stronger than Condition \ref{eq:linearcondition}, Condition \ref{condition:cond2} is satisfied when the linear instrumental variable model holds and the $p+q$-dimensional random vector $(\bX^\T, \bZ^\T)^\T$ follows an elliptical distribution. In our simulation study in Section \ref{sec:sims}, we show the proposed method can still exhibit reasonable performance even when this condition is not satisfied, for example, when $\bZ$ is binary. We then have the following result.

\begin{lemma}
Assume Condition \eqref{condition:cond2} holds. Then the quantity $ E(\bXhat \mid y)$ is contained in the linear subspace spanned by $\bSigma_{\bXhat}\bB$.
\label{lemma:SIR}
\end{lemma}

The covariance matrix $\bSigma_{\bXhat}$ in Lemma \ref{lemma:SIR} need not be of full-rank and invertible e.g., we permit the number of endogenous covariates $p$ to exceed the number of instruments $q$. 
To be precise, while theoretically we need to understand the behavior of its inverse, computationally $\bSigma_{\bXhat}$ does not have to be inverted in what follows.

Lemma \ref{lemma:SIR} motivates a two-stage lasso \SIR estimator for the column space of $\bB$ based on the observed vector $\by$, covariate matrix $\mcX$, and instrument matrix $\mcZ$. The two stages are detailed as follows: 
In {stage 1}, we identify and estimate the effects of the instruments $\bGamma$, and subsequently obtain the predicted values of the covariates $\bX$. To accomplish this, we fit a multivariate instrumental variable model given by the second equation in \eqref{eq:multipleindex_linearinstrument}, with the covariates $\bX$ as the $p$-dimensional response and instruments $\bZ$ as the $q$-dimensional predictor. Moreover, because we assume each endogenous covariate depends only on a few instrumental variables, 
then a sparse first stage estimator $\hat{\bGamma}$ is obtained by computing in parallel for $j = 1,\ldots,p$
\begin{align*}
\hat{\bgamma}_j = \arg\!\min_{\bgamma} \dfrac{1}{2n} \Vert \bx_j - \mcZ\bgamma  \Vert_F^2 + \mu_{1j} \Vert \bgamma\Vert_1, 
\end{align*}
where $\mu_{1j} > 0$ denotes the tuning parameter for the $j$th covariate in the first stage. Afterwards, the estimator $\hat{\bGamma} = [\hat{\bgamma}_1, \ldots, \hat{\bgamma}_p]$ is formed along with the fitted value matrix $\widehat{\mcX} = \mcZ\widehat{\bGamma}$.  

In {stage 2}, we compute a sparse estimator for the column space of $\bB$ by applying the lasso \SIR estimator \eqref{eq:SIR} with response vector $\by$ and the fitted value matrix $\hat{\mcX}$, 
Specifically, let $\widehat{\bLambda} = (nc)^{-1} \widehat{\mcX}^\T \bD \widehat{\mcX} = (nc)^{-1} \widehat{\bGamma}^\T {\mcZ}^\T \bD \mcZ\widehat{\bGamma}$, and define the pseudo-response vector $\tilde{\by}_k = (\hat\lambda_k)^{-1} \bD \widehat{\mcX} \hat{\boldeta}_k$, where $\hat{\boldeta}_k = \boldeta_k(\widehat{\bLambda})$ is the eigenvector associated with the $k$th largest eigenvalue $\hat\lambda_k = \lambda_k(\hat{\bLambda})$, and $\bD$ is defined as in Section \ref{subsection:review}. The two-stage lasso \SIR estimator for the $k$th column of $\bB$ then is defined as
\begin{equation}
\hat{\bbeta}_k = \arg\!\min_{\bbeta} \dfrac{1}{2n} \Vert \tilde{{\by}}_k - \widehat{\mcX}\bbeta\Vert_2^2 + \mu_{2k} \Vert \bbeta\Vert_1, \; k=1,\ldots, d,
\label{eq:2stagesmultipleindex}
\end{equation}
where $\mu_{2k} > 0$ is a tuning parameter for the $k$th pseudo-response in the second stage.

Computationally, both stages can be implemented straightforwardly using  coordinate descent algorithms e.g., we apply the \texttt{R} package \texttt{glmnet} \citep{friedman2010regularization} in both stages.

\section{Theoretical analysis} \label{section:theory}
We study the properties of the two-stage lasso \SIR estimator for the column space of ${\bB}$, under the multiple index model \eqref{eq:multipleindex_linearinstrument} and when the dimensionality of the covariates $p$ and instruments $q$ can grow exponentially with sample size $n$. 
Compared with the theoretical results for the linear outcome model in \citet{lin2015regularization}, the major challenge here lies in the nonlinear nature of the conditional expectation $E(\widehat{\bX} \mid y)$, and subsequently the behavior of the eigenvectors corresponding to the covariance matrix $\Lambda = \Cov\{E(\widehat{\bX} \mid y)\}$ which require more complex technical developments. 
In what follows, we will use $C$, $C^\prime$, $C^{\prime\prime}$ to denote generic positive constants. 

Following \citet[Chapter 7,][]{wainwright2019high}, we say an $n \times m$ matrix $\bA$ satisfies a restricted eigenvalue condition over a set $S$ with parameters $(\kappa, \nu)$ if
$
\Vert \bA^{1/2} \bdelta\Vert_2^2 \geq \kappa \Vert \bdelta\Vert_2^2,
$
for all $\bdelta \in \mathcal{C}(S, \nu) = \left\{\delta \in \mathbb{R}^{m}: ~\Vert \bdelta_{S^c} \Vert_1 \leq \nu \Vert \bdelta_S \Vert_1 \right\}$. Also, for a vector $\bv$ let $\supp(\bv) = \{j : v_j \neq 0\}$ denote the support set of $\bv$, and 
$\sigma_{\max} = \max_{1\leq j\leq p} \sigma_j$ where we recall $\sigma^2_j$ are the diagonal elements of the covariance matrix $\Sigma$ corresponding to $(\bU_i^\T, \varepsilon_i)^\T \in \mathbb{R}^{p+1}$. We assume each $\sigma_j^2$ and hence $\sigma_{\max}$ to be a constant with respect to $n$. 

The following technical conditions on the distribution of the $\bZ$ and the conditional expectation $E(\bX \mid \bZ) = \bGamma^\T \bZ$ are required.

\begin{condition}
The instrument random vector $\bZ$ is sub-Gaussian with a $q\times q$ covariance matrix $\bSigma_{\bZ}$, and for any $\bv \in \mathbb{R}^{q}$ there exists a constant $C$ such that $\norm{\bZ}_{\psi_2} \leq C$, where
$
\norm{\bZ}_{\psi_2} = \sup_{\bv \in \mathbb{R}^{q}, \norm{\bv} = 1} \norm{\bv^\T\bZ}_{\psi_2}, 
$
and $\norm{u}_{\psi_2} = \sup_{m} m^{-1/2} \left(E\vert u \vert^{m}\right)^{1/m}$.
\label{condition:subGaussian}
\end{condition}

\begin{condition}
There exist positive constants $L$ and $C$ such that $\Vert \bGamma \Vert_1 \leq L$, and the largest singular values $\Vert\bGamma \Vert_2 \leq C$. 
\label{condition:boundedGamma}
\end{condition}

\begin{condition}
The quantities $p,q,r,s$ satisfy $rs \left\{(\log pq)/n \right\}^{1/2} = o(1)$.
\label{condition:rate}
\end{condition}

The sub-Gaussian condition is common in high-dimensional data analysis \citep[e.g.][]{tan2018convex, wainwright2019high, nghiem2023screening}, with a direct consequence being that the instrumental matrix $\mcZ$ satisfies the \RE condition with parameter $(\kappa, 3) $ with high probability when $n > Cr\log q$, for some finite $\kappa$ \citep{estimation2014norm}; we will condition on this event in our development below.  Furthermore, letting $\widehat{\bSigma}_{\bZ} = \mcZ^\T\mcZ/n$ denote the sample covariance of $\mcZ$ data, then 
\begin{equation}
\|\widehat{\bSigma}_{\bZ}-\bSigma_{\bZ}\|_{\max }=C\sqrt{\frac{\log q}{n}},
\label{eq:estSigmaZ}
\end{equation}
with probability at least $1-\exp(-C^\prime \log q)$ \citep[see also][]{ravikumar2011high, tan2018convex}.  
Condition \ref{condition:boundedGamma} is also imposed in the theoretical analysis of the two-stage penalized least square methods in \citet{lin2015regularization}, and is necessary to bound the estimation error from the first stage. 
Finally, we allow the number of covariates $p$, instruments $q$, and the sparsity levels $s$ and $r$ to depend on the sample size $n$, subject to the rate of growth of these parameters dictated by Condition \ref{condition:rate}. For example, if $r$ and $s$ are bounded, then $p$ and $q$ can grow exponentially with $n$. 

\subsection{First stage results}
In the first stage of the two-stage lasso \SIR procedure, each column of the matrix $\bGamma$ is a lasso estimator from the regression of $\bx_j$ on the design matrix $\mcZ$. For completeness, we re-state the following result from \citet{lin2015regularization}, which itself is a direct application of \citet{estimation2014norm}

\begin{theorem}
\label{theorem:firststage}
Assume Conditions \ref{condition:subGaussian} and \ref{condition:boundedGamma} are satisfied, and set 
\begin{equation}
\mu_{1j} = C \sigma_j \sqrt{\dfrac{\log
 pq }{n}}, ~j=1,\ldots, p.
\label{eq:tuningparameter1ststage1}
\end{equation}
Then with probability at least $1-\exp(-C^\prime \log q)$, the columns of  $\widehat{\bGamma}$ satisfy 
$
\norm{\hat{\bgamma}_j - \bgamma_j}_{1} \leq C  \sigma_{\max}  r \{\log (pq)/n\}^{1/2},
$
and with probability at least $1-\exp(-C^\prime \log pq)$, we obtain
\begin{equation}
\left\|\widehat{{\bGamma}}-{\bGamma}\right\|_1 \leq  C \sigma_{\max} r\sqrt{\frac{\log pq}{n}}.
\label{eq:tuningparameter1ststage2}
\end{equation}
\end{theorem}

\noindent 
Theorem \ref{theorem:firststage} allows us to establish estimation consistency for the column space of $\bB$ in the following section. 

\subsection{Second stage results}
In the second stage, we compute the lasso \SIR estimator using the outcome $y$ and the fitted value matrix  $\widehat{\mcX} = \mcZ\widehat{\bGamma}$ obtained from the first stage. Since $\bX$ is related to the instruments $\bZ$ via \eqref{eq:multipleindex_linearinstrument}, then we have $E(\widehat{\bX} \mid y) = {\bGamma}^\T E(\bZ\mid y)$, where  $ E(\bZ\mid y)$ is related to the dimension reduction subspace $\mathcal{S}_{y \mid \bZ}$ of $y$ on $\bZ$. It follows that we can impose technical conditions similar to those in \citet{lin2019sparse} to ensure this subspace can be estimated consistently in high-dimensional settings:

\begin{condition}
There exist constants $C_{\min}$ and $C_{\max}$ such that $0 < C_{\min} \leq \lambda_q(\bSigma_{\bZ})  \leq \cdots \leq  \lambda_{1}(\bSigma_{\bZ}) < C_{\max} < \infty$. 
\label{condition:restrictedeigenvector}
\end{condition}

\begin{condition} 
There exist positive constants $c$ and $\kappa \geq 1$ such that $0 < c < \lambda_d(\bOmega) \leq \ldots \leq \lambda_1(\bOmega) \leq \kappa\lambda_d(\bOmega) < \lambda_1(\bSigma_{\bZ})$, where $\bOmega = \Cov\left\{E\left({\bZ} \mid y\right)\right\}$, and $n\lambda_1(\bOmega) = q^\alpha$ for some $\alpha >1/2$.
\label{condition:lowerboundsmallest}
\end{condition}

\begin{condition}
The quantity ${m}_{\bZ}(y) = E(\bZ \mid y)$ satisfies the sliced stability condition introduced in \citet{lin2018consistency}.
\label{condition:partition}
\end{condition}

\noindent Condition \ref{condition:restrictedeigenvector} ensures the loss function corresponding to the second stage of the proposed lasso \SIR estimator does not have too small curvature, and is similar to a condition imposed by \citet{lin2015regularization} for the two-stage penalized least square method. 
Condition \ref{condition:lowerboundsmallest} is analogous to the coverage condition imposed in the \SIR literature for $\bLambda$ \citep[e.g.,][]{li1991sliced}.   
Finally, Condition \ref{condition:partition} controls the smoothness of the conditional expectation ${m}_{\bZ}(y) = E(\bZ \mid y)$. Such a condition is similarly used in \citet{tan2018convex} when no endogeneity is present, with analogous conditions on the total variation having been employed earlier by \citet{hsing1992asymptotic, zhu2006sliced} for other sufficient dimension reduction methods.

We next establish the consistency of the two-stage lasso \SIR estimator for $\text{span}(\bB)$.
First, consider the case of $d=1$ i.e., the single index model version of \eqref{eq:multipleindex_linearinstrument}. Here, Lemma \ref{lemma:SIR} implies the covariance matrix $\bLambda = \Cov\{E(\bXhat \mid y) \}$ has rank $d=1$ and hence can be written in the form $\bLambda = \lambda \boldeta \boldeta^\T$, where $\lambda > 0$ denotes the non-zero eigenvalue of $\bLambda$ and $\boldeta$ is the corresponding eigenvector. Without loss of generality, assume $\Vert \bbeta \Vert = 1$.  For convenience below, we omit the subscript $k$ and use $\hat\lambda$ and $\hat\boldeta$ to denote the largest eigenvalue and its corresponding eigenvector of $\hat\bLambda$, respectively, noting both were defined above \eqref{eq:2stagesmultipleindex}. Furthermore, we write $\tilde{\by}_1 = \tilde\by$. With this notation, the two-stage lasso \SIR estimator for the column space of $\bbeta$ 
is given by
\begin{equation}
\hat\bbeta = \argmin_{\bbeta \in \mathbb{R}^{p}} \left\{ \dfrac{1}{2n} \Vert \tilde{\by}- \widehat{\mcX}\bbeta \Vert _2^2 + \mu_2 \Vert \bbeta \Vert_1 \right\}.
\label{eq:solutionlasso}
\end{equation}
By Lemma \eqref{lemma:SIR}, we have $\bSigma_{\bXhat}\bbeta = \xi \boldeta$, where $\xi = \norm{\bSigma_{\bXhat}\bbeta}_2 > 0$ is a constant. Letting $\tilde{\boldeta}  = P(\boldeta)\hat{\boldeta} = \boldeta\boldeta^\T\hat{\boldeta} = \bSigma_{\bXhat}\tilde{\bbeta}$, where $\tilde{\bbeta} = \xi^{-1} \bbeta{\boldeta}^\T \hat{\boldeta}$, then it is straightforward to see $\tilde\bbeta$ has the same direction as $\bbeta$, i.e., $\mathcal{P}(\tilde\bbeta) = \mathcal{P}(\bbeta)$, when $\boldeta^\T\hat{\boldeta} \neq 0$, which occurs with probability one. With this quantity, we obtain the following deterministic result.

\begin{lemma}
\label{lemma:deterministic}
Suppose the sample covariance matrix $\widehat{\bSigma}_{\bXhat} = n^{-1} \widehat{\mcX}^\T \widehat{\mcX}$ satisfies an \RE property with parameters $(\kappa_2, 3)$; see Proposition 11 in Appendix B that shows the latter holds with high probability. If the tuning parameter satisfies $\mu_2 \geq 2 \left\Vert \hat\boldeta - n^{-1} \hat{\mcX}^{\T} \hat{\mcX}\tilde\bbeta \right\Vert_\infty$, then any solution $\hat\bbeta$ defined in \eqref{eq:solutionlasso} satisfies
\begin{align*}
\norm{\hat{ \bbeta}-\tilde{ \bbeta}}_2  \leq \dfrac{12}{\kappa_2}\mu_2 \sqrt{s}.
\end{align*}
\end{lemma}

Following Lemma \ref{lemma:deterministic}, by selecting a value of  $\mu_2$ so that the condition $\mu_2 \geq 2 \left\Vert \hat\boldeta - n^{-1} \hat{\mcX}^{\T} \hat{\mcX}\tilde\bbeta \right\Vert_\infty$ is satisfied with high probability, we obtain the consistency of the two-stage lasso \SIR estimator for the single index model. 

\begin{theorem}
\label{theorem:singleindex}
Assume Conditions \ref{condition:subGaussian}-\ref{condition:partition} are satisfied, and choose the first stage tuning parameters $\mu_{1j}$ as in \eqref{eq:tuningparameter1ststage1}. Let $\boldeta$ be the eigenvector associated with the largest eigenvalue of $\bm\Omega = \text{Cov}\{E(\bZ \mid y)\}$. Let $\rho_1= L^2 + \lambda^{-1/2}{L(1+ \Psi_1)}$ and $\rho_2 = L \sigma_{\max} + \lambda^{-1/2}(1+\Psi_1)\Psi_2$, where $\Psi_1 = \left\Vert \mathbf{P} \right\Vert_1$ and $\bP$ is the projection matrix associated with the space spanned by $\boldeta\bGamma$ and $\Psi_2 = \Vert \boldeta\boldeta^\top \Vert_1$. If we select the second stage tuning parameters as
\begin{equation}
\mu_2 = C \left(\rho_1 \sqrt{\frac{\log q}{n}} + \rho_2 r\sqrt{\frac{\log pq}{n}} \right),
\label{eq:mu2}
\end{equation}
then 
with probability at least $1- \exp(-C^\prime \log q) - \exp(-C^{\prime\prime} \log pq)$, the two-stage lasso \SIR estimator $\hat{\bbeta}$ satisfies 
\begin{align*}
\left\|P(\hat\bbeta) - P({\bbeta})\right\|_F \leq C\frac{\sqrt{s}}{\kappa_2}\left(\rho_1 \sqrt{\frac{\log q}{n}} + \rho_2 r\sqrt{\frac{\log pq}{n}} \right).
\end{align*}
\end{theorem}

Compared with the bound for the one-stage lasso \SIR estimator, which is established by \citet{lin2019sparse}, the theoretical bound for the two-stage lasso \SIR estimator has an additional factor of order $\left\{\log(pq)/n\right\}^{1/2}$, reflecting the consequences of endogeneity. Also, compared with the result for the two-stage penalized least squares estimator in the linear outcome model of \citet{lin2015regularization}, the bound in Theorem \ref{theorem:singleindex} involves an additional term $\rho_1\left(\log q/n \right)^{1/2}$, which 
reflects the consequences of non-linearity. 
Intuitively, the estimation error of $\hat{\bbeta}$ depends on the estimation error of $\bSigma_{\widehat{\bX}} = \bGamma^\T\bSigma_{\bZ} \bGamma$, which in turn depends upon the estimation error of both the coefficients $\hat{\bGamma}$ in the first stage and the covariance matrix $\bSigma_{\bZ}$. 

Next, we establish the selection consistency of the two-stage lasso \SIR estimator. 
Let $\hat{S} = \{j: \hat{\beta}_j \neq 0\}$, and partition the matrix $\bSigma_{\bXhat} = \bGamma^\T \bSigma_{\bZ} \bGamma$ as 
\begin{align*}
\bSigma_{\bXhat} = \begin{bmatrix}
\bC_{SS} & \bC_{S^c S} \\
\bC_{SS^c} & \bC_{S^c S^c}
\end{bmatrix}.
\end{align*}
Let $\phi = \Vert (\bC_{SS})^{-1} \Vert_\infty$, and note in this case we have $\xi = \norm{\bSigma_{\bXhat} \bbeta}_2 = \norm{\bC_{SS}\bbeta_S}_2$. We require an additional condition as follows. 

\begin{condition}
There exists a constant $0<\alpha\leq 1$ such that $\Vert \bC_{S^cS} (\bC_{SS})^{-1} \Vert_\infty \leq 1-\alpha$.
\label{condition:irreprentability}
\end{condition}

Condition \ref{condition:irreprentability} is similar to the mutual incoherence condition for the ordinary lasso estimator in the linear model without endogeneity \citep[e.g., see Section 7.5.1,][]{wainwright2019high}.

\begin{theorem}
\label{theorem:singleindex_variableselection}
Assume Conditions \ref{condition:subGaussian}-\ref{condition:irreprentability} are satisfied, and choose the first stage tuning parameters $\mu_{1j}$ as in \eqref{eq:tuningparameter1ststage1}. If there exists a constant $C$ such that
\begin{equation}
CsL\left\{L\sqrt{\dfrac{ \log q}{n}} +   \sigma_{\max} r \sqrt{\dfrac{\log pq}{n}} \right\} \leq \frac{\alpha}{\phi(4-\alpha)},
\label{eq:add.constraint}
\end{equation}
and the second stage tuning parameter is selected to satisfy
\begin{equation}
\min_{j \in S} \vert \beta_{j} \vert >\frac{2\xi\boldeta^\T \widehat{\boldeta}}{2-\alpha} \varphi \mu_2,
\label{eq:minimumsignal}
\end{equation}
then 
with probability at least $1- \exp(C^\prime \log q) - \exp(C^{\prime \prime} \log pq)$, the support of the two-stage lasso \SIR estimator $\hat{\bbeta}$ satisfies $\hat{S} = S$.
\end{theorem}

Since the mutual incoherence condition \ref{condition:irreprentability} is imposed for the population quantity $\bSigma_{\bXhat}$, the additional assumption in \eqref{eq:add.constraint} is necessary to ensure that this condition is satisfied by the sample covariance matrix $\widehat{\bSigma}_{\bXhat}$ with high probability. 
Also, \eqref{eq:minimumsignal}
has an additional factor $\xi\boldeta^\T \widehat{\boldeta}$ in the numerator, 
reflecting another consequence of non-linearity. 
On the other hand, the lower bound is small when $\alpha$ is small i.e., the lower the correlation between the important and non-important variables, the smaller the magnitude of non-zero components which can be identified.

\subsection{Extension to multiple index models}
We generalize the above results to multiple index models. For $k = 1,\ldots,d$, let $\widehat{{S}}_k = \{j: \vert \hat{\beta}_{jk} \vert \neq 0\}$ denote the estimated support set for the vector $\hat{\bbeta}_k$, and $\widehat{{S}} = \cup_k{\mathcal{S}_k}$. Recall $\bB$ is only identifiable up to rotational transformation, and the latter only preserves the zero rows of $\bB$. As such, selection in the multiple index model refers only to selecting non-zero rows of $\bB$. Equivalently, the set $\hat{{S}}$ should be a consistent estimator of the support set  $S$ for the diagonal elements of $\mathcal{P}(\bB)$.

\begin{theorem}
Assume Conditions \ref{condition:subGaussian}-\ref{condition:partition} are satisfied, and choose the first stage tuning parameters $\mu_{1j}$ as in \eqref{eq:tuningparameter1ststage1}. Let $\rho_1= L^2 + \lambda_d^{-1/2}L(1+ \Psi_1)$ and $\rho_2 = L \sigma_{\max} + \lambda_d^{-1/2}(1+\Psi_1)\Psi_2$, where $\lambda_d$ is the $d$th largest eigenvalue of $\Lambda = \Cov\{E(\bXhat \mid y) \}$. If we select the second stage tuning parameters to satisfy
\begin{equation}
\mu_{2k} = C \left(\rho_1 \sqrt{\frac{\log q}{n}} + \rho_2 r\sqrt{\frac{\log pq}{n}} \right), \; k=1,\ldots,d,
\label{eq:mu2k}
\end{equation}
\begin{enumerate}
    \item[(a)] Then with probability at least $1- d\exp(-C^\prime \log q) - d\exp(-C^{\prime\prime} \log pq)$, the two-stage lasso \SIR estimator $\widehat{\bB} =(\hat\bbeta_1, \ldots, \hat\bbeta_d)$, whose columns $\hat{\bbeta}_k$ are defined by \eqref{eq:2stagesmultipleindex}, satisfies 
\begin{align*}
\left\|P(\widehat{\bB}) - P({\bB})\right\|_F \leq C\frac{\sqrt{s}}{\kappa_2}\left(\rho_1 \sqrt{\frac{\log q}{n}} + \rho_2 r\sqrt{\frac{\log pq}{n}} \right).
\end{align*}

\item[(b)] If Condition \ref{condition:irreprentability} also holds and at least one tuning parameter $\mu_{2k}$ satisfies
\begin{align*}
\min_{j \in \mathcal{S}^k} \vert \beta_{jk}  \vert >\frac{2\xi_k{\boldeta}_k^\T \widehat{\boldeta}_k}{2-\alpha} \varphi \mu_{2k},
\end{align*}
 where $\xi_k = \Vert \Sigma_{\bXhat} \bbeta_k \Vert_2$, then $\widehat{\mathcal{S}} = \mathcal{S}$ with the same probability,.
\end{enumerate}
\label{theorem:multipleindexmodel}
\end{theorem}

Parts (a) and (b) in Theorem \ref{theorem:multipleindexmodel} for multiple index models essentially require Theorems \ref{theorem:singleindex} and \ref{theorem:singleindex_variableselection}, respectively, to hold for every dimension. As such, the upper bound on the estimation error of $\mathcal{P}(\widehat{\bB})$ depends on the lowest non-zero eigenvalue $\lambda_d$ of $\bLambda$, and requires all sample eigenvectors $\widehat{\boldeta}_k$ to not be orthogonal to their corresponding eigenvector $\boldeta_k$ for $k=1,\ldots, d$. Under these conditions, we can establish bounds for the estimation and selection consistency for the multiple index model \eqref{eq:multipleindex_linearinstrument} under endogeneity.

\section{Simulation studies} \label{sec:sims}
We conducted a numerical study to evaluate the performance of the proposed two-stage lasso \SIR estimator for recovering the (sparsity patterns of the) dimension reduction subspace.
We simulated data $\{(y_i, \bX_i, \bZ_i); i=1,\ldots,n\}$, from a linear instrumental variable model $\bX_i = \bGamma^\T \bZ_i + \bU_i$ coupled with one of the following five models for the response:
\begin{align*}
\text{(i)}: \; y_i & = \bX_i^\T \bbeta_1 + \varepsilon_i, ~~  
\text{(ii)}: \; y_i = \exp(\bX_i^\T \bbeta_1 + \varepsilon_i), \quad \text{(iii)}: \; y_i = \sinh(\bX_i^\T \bbeta_1 + \varepsilon_i)\\
\text{(iv)}: \; y_i &= (\bX_i^\T\bbeta_2) \exp(\bX_i^\T \bbeta_1 + \varepsilon_i), \quad \text{(v)}: \; y_i = \frac{\exp(\bX_i^\T \bbeta_1 + \varepsilon_i)}{3/2 + \bX_i^\T \bbeta_2 + \varepsilon_i}.
\end{align*}
Model (i) is the linear model, while the single and double index models (ii) - (v) are similar to those previously used in the literature to assess SDR methods \citep[e.g.,][]{lin2019sparse}. The true sparse matrix $\bB = \bbeta_1$ in the single index models (i) and (ii),  and $\bB = [\bbeta_1, \bbeta_2]$ in the multiple index models (iii) and (iv) are constructed as follows: We randomly selected $s=5$ rows of $\bB$ to be nonzero and generated each non-zero entry in these rows from a uniform distribution on the two disjoint intervals $[-1,-0.5]$ and $[0.5,1]$, which we denote as $U([-1, -0.5] \cup [0.5, 1])$. The elements in all other rows were set to zero. We then used a Gram-Schmidt process to ensure $\bbeta_1$ and $\bbeta_2$ were orthogonal to each other in the case of the double index model. The covariate $\bX_i$ vector was generated from the instrumental variable model $\bX_i = \bGamma^\T \bZ_i + \bU_i$ for $i=1,\ldots, n$, where we followed a similar generation mechanism to that in \citet{lin2015regularization}. That is, we simulated each of the $q$ elements of $\bZ_i$ from either: (1) a standard normal distribution, or (2) a Bernoulli distribution with success probability $0.5$. To generate $\bGamma$, we randomly set $r=5$ nonzero entries in each column from the uniform distribution $U([-b,-a] \cup[a, b])$ with $(a,b) = (0.75, 1)$. Finally, we generated the random noise vector $(\bU_i^\T, \varepsilon_i)^\T$ from a $(p+1)$-variate Gaussian distribution with zero mean vector and covariance matrix $\bSigma =(\sigma_{ij})$ as follows: We set $\sigma_{i j}=(0.2)^{|i-j|}$ for $i, j=$ $1, \ldots, p$, so that $\bSigma_{\bU}$ has an AR(1) structure. For the last column/row of $\bSigma$ characterizing endogeneity, we set $\bSigma_{S, p+1} = -\bSigma_{S,S} \bbeta_{1S}$, where $S$ denotes the index set of non-zero elements in $\bbeta_1$, and randomly selected another five elements and set them equal to $0.3$, while the remaining elements were set to zero. The last element $\sigma_{p+1, p+1} = \bSigma_{\bU, \varepsilon}^\T \bSigma_{\bU,\bU} \bSigma_{\bU, \varepsilon} + U(0, 0.2)$. This construction of $\bSigma$ ensures it is positive definite, and that $\bSigma_{\bX}^{-1}\bSigma_{\bX, \epsilon}$ does not belong to the column space of $\bB$. 

We considered both low-dimensional and high-dimensional settings, where for the former we set $p=q=40$ and varied $n \in \{200, 500\}$, while for high-dimensional settings we fixed $n=200$ and varied $p=q \in \{500, 1000\}$. For each configuration, we simulated $100$ datasets, where in each dataset we applied the one-stage lasso, lasso \SIR, two-stage lasso, and the proposed two-stage lasso \SIR estimators.
Note both the one-stage lasso and two-stage lasso estimates assume $d=1$, while the dimension $d$ is assumed to be known for both the lasso \SIR and two-stage lasso \SIR. For the latter two \SIR-based estimators, we also set the number of slices $H=10$. 
To reduce the computational cost, for the two-stage lasso and two-stage lasso \SIR estimators, we selected the tuning parameter for the penalized regression of each covariate on instruments to minimize the Bayesian information criterion (\textsc{BIC}). We also found that using other information criterion, such as extended BIC of \citet{chen2008extended}, led to similar results in the second stage in both low and high-dimensional settings.     
All other tuning parameters were selected via an appropriate ten-fold cross-validation approach. We assessed performance of all methods based on the error on the estimated projection matrix $\Vert \mathcal{P}(\hat{\bB}) - \mathcal{P}(\bB)\Vert_F$, along with the area under the receiver operating curve (\textsc{AUC}) score for variable selection performance.

Table \ref{tab:simulation_lowdimension} demonstrates that in low-dimensional settings, the proposed two-stage lasso \SIR estimator performed best in terms of both estimation error and variable selection. Indeed, it is the only method among the four considered whose estimation error consistently decreased when sample size increased from $n=200$ to $n=500$. The one-stage lasso estimator, which ignores both endogeneity and non-linearity in the outcome model, performed the worst as expected. Comparing the lasso \SIR and two-stage lasso, the former tended to perform better except in model (i), suggesting the consequence of ignoring non-linearity and/or underestimating the dimensions were more serious than ignoring endogeneity in this simulation design. 
Turning to the variable selection results, the lasso \SIR estimator has overall good performance, while the two-stage lasso estimator only achieves an AUC close to $1$ in the linear model (i).  The proposed two-stage lasso \SIR estimator consistently had the (equal) best selection performance. 

For high-dimensional settings, Table \ref{tab:simulation_highdimension} shows that the proposed two-stage lasso \SIR still had the best performance overall. For the linear model (i), the two-stage lasso performed well as expected, but its performance was worse than the two-stage lasso \SIR estimator in single index models (ii)-(iii) where the link function is non-linear. The proposed two-stage lasso \SIR also had a much smaller estimation error compared with the one-stage lasso \SIR in all single index models (i)--(iii), although the two methods had quite similar estimation error in the double index models (iv)-(v).
Finally, the performance of all methods in both low- and high-dimensional settings was similar for both continuous and binary cases of $\bZ$. In particular, this suggested the proposed two-stage lasso \SIR estimator was not overly sensitive to violation of Condition \ref{condition:cond2} in this simulation design.

\begin{table}[t]
\centering

\scalebox{0.73}{
\begin{tabular}[t]{llr lll>{\bfseries}l lll>{\bfseries}l}
\toprule[1.5pt]
$\bZ$ & Model & $n$ & \multicolumn{4}{c}{Error} & \multicolumn{4}{c}{AUC} \\
\cmidrule(lr){4-7}  \cmidrule(lr){8-11}
& & & lasso & LSIR & 2Slasso & \normalfont 2SLSIR & lasso & LSIR & 2Slasso &  \normalfont 2SLSIR  \\
\midrule
Cont. & (i) & 200 & 0.62 (0.16) & 0.47 (0.13) & 0.25 (0.15) & 0.18 (0.07) & 0.99 (0.02) & 0.99 (0.03) & \bf 1.00 (0.02) & 1.00 (0.00)\\
 &  & 500 & 0.65 (0.13) & 0.44 (0.13) & 0.14 (0.07) & 0.09 (0.03) & 0.99 (0.02) & \bf 1.00 (0.01) & \bf 1.00 (0.00) & 1.00 (0.00)\\
 & (ii) & 200 & 1.13 (0.17) & 0.47 (0.13) & 1.03 (0.20) & 0.18 (0.07) & 0.73 (0.12) & 0.99 (0.03) & 0.77 (0.13) & \bf 1.00 (0.00)\\
 &  & 500 & 1.13 (0.14) & 0.44 (0.13) & 1.02 (0.17) & 0.09 (0.03) & 0.72 (0.12) & 1.00 (0.01) & 0.77 (0.14) & 1.00 (0.00)\\
 & (iii) & 200 & 1.10 (0.19) & 0.47 (0.13) & 0.95 (0.25) & 0.18 (0.07) & 0.73 (0.14) & 0.99 (0.03) & 0.81 (0.15) & \bf 1.00 (0.00)\\
 &  & 500 & 1.05 (0.18) & 0.44 (0.13) & 0.90 (0.23) & 0.09 (0.03) & 0.78 (0.13) & \bf 1.00 (0.01) & 0.83 (0.14) & 1.00 (0.00)\\
 & (iv) & 200 & 1.48 (0.24) & 0.87 (0.19) & 1.45 (0.24) & 0.67 (0.22) & 0.60 (0.10) & \bf 1.00 (0.01) & 0.65 (0.09) & 1.00 (0.01)\\
 &  & 500 & 1.44 (0.22) & 0.81 (0.21) & 1.36 (0.22) & 0.39 (0.17) & 0.62 (0.09) & \bf 1.00 (0.01) & 0.65 (0.13) & 1.00 (0.00)\\
 & (v) & 200 & 1.63 (0.12) & 0.85 (0.21) & 1.57 (0.15) & 0.70 (0.21) & 0.56 (0.07) & 0.99 (0.02) & 0.60 (0.11) & 1.00 (0.01)\\
 &  & 500 & 1.60 (0.19) & 0.75 (0.22) & 1.51 (0.23) & 0.41 (0.19) & 0.53 (0.06) & \bf 1.00 (0.01) & 0.60 (0.07) & 1.00 (0.02)\\
 \addlinespace 
Bin & (i) & 200 & 0.61 (0.16) & 0.46 (0.15) & 0.25 (0.18) & 0.17 (0.06) & 0.98 (0.03) & 0.99 (0.03) & 1.00 (0.02) & 1.00 (0.00)\\
 &  & 500 & 0.68 (0.14) & 0.48 (0.15) & 0.14 (0.08) & 0.09 (0.03) & 0.99 (0.03) & 0.99 (0.02) & 1.00 (0.00) & 1.00 (0.00)\\
 & (ii) & 200 & 1.09 (0.18) & 0.46 (0.15) & 1.04 (0.21) & 0.17 (0.06) & 0.73 (0.11) & 0.99 (0.03) & 0.77 (0.12) & 1.00 (0.00)\\
 &  & 500 & 1.03 (0.21) & 0.48 (0.15) & 0.94 (0.22) & 0.09 (0.03) & 0.79 (0.13) & 0.99 (0.02) & 0.81 (0.13) & 1.00 (0.00)\\
 & (iii) & 200 & 1.07 (0.17) & 0.61 (0.17) & 0.94 (0.23) & 0.22 (0.12) & 0.76 (0.12) & 0.98 (0.03) & 0.81 (0.13) & 1.00 (0.01)\\
 &  & 500 & 0.99 (0.19) & 0.67 (0.14) & 0.82 (0.24) & 0.13 (0.05) & 0.82 (0.12) & 0.98 (0.03) & 0.88 (0.13) & 1.00 (0.00)\\
 & (iv) & 200 & 1.47 (0.22) & 0.80 (0.22) & 1.41 (0.19) & 0.61 (0.22) & 0.62 (0.09) & \bf 1.00 (0.01) & 0.62 (0.07) & 1.00 (0.01)\\
 &  & 500 & 1.35 (0.21) & 0.80 (0.22) & 1.35 (0.20) & 0.38 (0.17) & 0.64 (0.07) & \bf 1.00 (0.01) & 0.66 (0.09) & 1.00 (0.00)\\
 & (v) & 200 & 1.53 (0.18) & 0.82 (0.20) & 1.55 (0.15) & 0.64 (0.20) & 0.61 (0.10) & \bf 1.00 (0.01) & 0.63 (0.12) & 1.00 (0.01)\\
 &  & 500 & 1.46 (0.23) & 0.81 (0.22) & 1.52 (0.19) & 0.39 (0.16) & 0.66 (0.14) & \bf 1.00 (0.02) & 0.68 (0.16) & 1.00 (0.00)\\
\bottomrule[1.5pt]
\end{tabular}}
\caption{Simulation results for the lasso, one-stage lasso \SIR (LSIR), two-stage lasso (2Slasso), and the two-stage lasso \SIR (2SLSIR) estimators, in low-dimensional settings with $p=q=40$. Standard errors are included in parentheses. The lowest estimation error and the highest AUC score are highlighted in each row.} 
\label{tab:simulation_lowdimension}
\end{table}

\begin{table}[t]
\centering
\scalebox{0.73}{
\begin{tabular}[t]{llr lll>{\bfseries}l lll>{\bfseries}l}
\toprule[1.5pt]
$\bZ$ & Model & $p=q$ & \multicolumn{4}{c}{Error} & \multicolumn{4}{c}{AUC} \\
\cmidrule(lr){4-7}  \cmidrule(lr){8-11}
& & & lasso & LSIR & 2Slasso &  \normalfont 2SLSIR & lasso & LSIR & 2Slasso &  \normalfont 2SLSIR  \\
\midrule
  \addlinespace
Cont. & (i) & 500 & 0.31 (0.09) & 0.41 (0.14) & \bf 0.13 (0.03) & \normalfont 0.25 (0.12) & \bf 1.00 (0.00) & 0.97 (0.05) & \bf 1.00 (0.00) & \normalfont 0.98 (0.04)\\
 &  & 1000 & 0.31 (0.10) & 0.61 (0.16) & \bf 0.13 (0.04) & \normalfont 0.35 (0.13) & \bf 1.00 (0.00) & 0.97 (0.05) & \bf 1.00 (0.00) & \normalfont 0.98 (0.04)\\
 & (ii) & 500 & 1.27 (0.13) & 0.39 (0.13) & 1.22 (0.14) & 0.21 (0.10) & 0.66 (0.12) & \bf 0.99 (0.03) & 0.72 (0.13) & 0.99 (0.03)\\
 &  & 1000 & 1.30 (0.14) & 0.63 (0.16) & 1.25 (0.14) & 0.38 (0.14) & 0.57 (0.06) & \bf 0.97 (0.05) & 0.60 (0.07) & 0.97 (0.05)\\
 & (iii) & 500 & 1.16 (0.21) & 0.39 (0.13) & 1.11 (0.18) & 0.21 (0.10) & 0.70 (0.13) & 0.98 (0.04) & 0.73 (0.14) & 0.99 (0.03)\\
 &  & 1000 & 1.20 (0.15) & 0.50 (0.18) & 1.18 (0.17) & 0.30 (0.15) & 0.68 (0.12) & 0.97 (0.06) & 0.68 (0.13) & 0.98 (0.04)\\
 & (iv) & 500 & 1.65 (0.12) & 1.05 (0.24) & 1.64 (0.18) & 1.00 (0.28) & 0.55 (0.05) & 0.99 (0.02) & 0.53 (0.05) & 1.00 (0.02)\\
 &  & 1000 & 1.61 (0.22) & \bf 0.95 (0.26) & 1.64 (0.17) & 0.95 (0.33) & 0.53 (0.05) & \bf 1.00 (0.01) & 0.53 (0.05) & 1.00 (0.02)\\
 & (v) & 500 & 1.68 (0.08) & 1.08 (0.25) & 1.66 (0.11) & 1.07 (0.28) & 0.54 (0.07) & \bf 1.00 (0.02) & 0.57 (0.10) & 1.00 (0.02)\\
 &  & 1000 & 1.69 (0.09) & 1.02 (0.28) & 1.73 (0.00) & 1.02 (0.30) & 0.52 (0.05) & \bf 0.99 (0.03) & 0.50 (0.00) & 0.99 (0.04)\\
 \addlinespace
Bin & (i) & 500 & 0.32 (0.09) & 0.39 (0.15) & \bf 0.13 (0.04) & \normalfont 0.23 (0.11) & \bf 1.00 (0.00) & 0.98 (0.05) & \bf 1.00 (0.00) & \normalfont 0.99 (0.04)\\
 &  & 1000 & 0.31 (0.10) & 0.47 (0.19) & \bf 0.13 (0.04) & \normalfont 0.28 (0.13) & \bf 1.00 (0.00) & 0.97 (0.07) & \bf 1.00 (0.00) & \normalfont 0.98 (0.04)\\
 & (ii) & 500 & 1.22 (0.19) & 0.42 (0.13) & 1.17 (0.16) & 0.21 (0.09) & 0.67 (0.13) & 0.98 (0.04) & 0.70 (0.11) & 0.99 (0.03)\\
 &  & 1000 & 1.26 (0.14) & 0.45 (0.14) & 1.23 (0.12) & 0.29 (0.17) & 0.62 (0.10) & \bf 0.97 (0.05) & 0.68 (0.12) & 0.97 (0.05)\\
 & (iii) & 500 & 1.18 (0.16) & 0.41 (0.14) & 1.07 (0.18) & 0.23 (0.11) & 0.70 (0.12) & 0.98 (0.04) & 0.77 (0.13) & 0.99 (0.03)\\
 &  & 1000 & 1.19 (0.17) & 0.48 (0.16) & 1.13 (0.17) & 0.30 (0.14) & 0.70 (0.12) & \bf 0.97 (0.05) & 0.74 (0.14) & 0.97 (0.05)\\
 & (iv) & 500 & 1.63 (0.19) & 1.03 (0.24) & 1.59 (0.16) & 0.98 (0.24) & 0.55 (0.08) & \bf 1.00 (0.02) & 0.57 (0.08) & 1.00 (0.01)\\
 &  & 1000 & 1.61 (0.17) & 0.94 (0.29) & 1.67 (0.13) & 0.94 (0.31) & 0.54 (0.05) & \bf 0.99 (0.04) & 0.53 (0.05) & 0.99 (0.03)\\
 & (v) & 500 & 1.63 (0.21) & 1.08 (0.23) & 1.65 (0.15) & 1.05 (0.26) & 0.56 (0.12) & \bf 0.99 (0.02) & 0.62 (0.14) & 1.00 (0.01)\\
 &  & 1000 & 1.71 (0.04) & 1.06 (0.28) & 1.71 (0.06) & 1.05 (0.28) & 0.52 (0.04) & \bf 0.99 (0.04) & 0.54 (0.10) & 0.99 (0.03)\\
\bottomrule[1.5pt]
\end{tabular}}
\caption{Simulation results for the lasso, lasso \SIR (LSIR), the two-stage lasso (2Slasso), and the two-stage lasso SIR (2SLSIR) estimators, in high-dimensional settings with $n=200$. Standard errors are included in parentheses. The lowest estimation error and the highest AUC score are highlighted in each row.} 
\label{tab:simulation_highdimension}

\end{table}

\section{Selection of dimensions}
\label{sec:selectiond}
An important problem in applying any sufficient dimension reduction method in practice is to estimate the dimension $d$ for the dimension reduction subspace. When there is no endogeneity, this can be estimated by examining the eigenvalues of $\widehat{\bLambda}^*$, which is an estimate of $\bLambda^* = \Cov\{E(\bX \mid y)\}$. In classical settings with $p$ fixed and $n \to \infty$, it can be shown that the $d$ largest estimated eigenvalues of $\widehat{\bLambda}^*$ are separated from the last $H-d$ eigenvalues, and hence we can estimate $d$ via an asymptotic test \citep{shao2007marginal} or a permutation/bootstrap test \citep[e.g.,][]{ye2003using}. The problem is more challenging in high-dimensional settings where $p/n \rightarrow 0$, since all non-zero eigenvalues of $\widehat{\bLambda}^*$ are of the same order $p/n$ and no longer separated. 
To overcome this issue, \citet{lin2019sparse} proposed selecting the structural dimension for the lasso \SIR estimator by clustering the adjusted eigenvalues $\tilde{\lambda}^*_k = \hat{\lambda}_k^* \Vert \hat{\bbeta}_k^* \Vert_2$, where $\hat{\bbeta}_k^*$ is a solution of \eqref{eq:SIR} for $k=1,\ldots, H$. 
Other sparse \SIR methods, such as those proposed by \citet{tan2018convex, zeng2024subspace}, select the structural dimension $d$ by augmenting an objective function with a nuclear norm constraint.

In our setting where endogeneity is present, the selection of dimensions in high dimensions is made more challenging by the complex eigenstructure of $\bLambda = \Cov\{E(\bXhat \mid y) \} = \bGamma^\T \Cov\{E(\bZ \mid y) \} \bGamma$. A first natural solution to this problem is to apply the clustering method in \citet{lin2019sparse} in Stage 2 of the proposed two-stage estimator, when we compute the lasso \SIR estimator of the outcome data $\by$ on the first stage fitted value matrix $\widehat{\mcX} = \mcZ \widehat{\bGamma}$.
On the other hand, 
we can also write  \eqref{eq:multipleindex_linearinstrument} as $y_i = f(\bZ_i^\T \bC, \tilde{\varepsilon}_i)$, where $\bC = \bGamma \bB$ is a $q \times d$ matrix. It follows that the central subspace of $y$ on $\bZ$ also has dimension $d$, assuming $\bC$ is full rank. Therefore, we also consider selecting $d$ simply by computing a lasso \SIR estimator of the outcome $y$ on the instrumental variables $\bZ$. 

We conducted a small simulation study to compare the performance of different methods to estimate $d$ based on applying the lasso \SIR estimator of $y$ on either $\mcZ$, $\mcX$, or $\widehat{\mcX}$. Specifically, after computing the lasso estimator on each dimension $k=1,\ldots,d$, we computed the adjusted eigenvalues as in \citet{lin2019sparse} and applied $K$-means clustering with $K=2$ on these adjusted eigenvalues. Finally, we estimated $\hat{d}$ as the size of the cluster containing the largest adjusted eigenvalues. To reduce the sensitivity of the final results to the choice of other tuning parameters, we performed $50$ repeated $5$-fold cross-validation (i.e. 50 random splits of the data) where within each repeat, we obtained the lasso \SIR estimator for each dimension at the tuning parameter selected as in the simulation, performed the clustering, and recorded the estimated $d$. The final estimator is then recorded as the value of $d$ selected with the highest frequency across 50 repeats.
We simulated data using the same setup as in Section \ref{sec:sims}, but only considered settings with continuous instruments $\bZ$ for brevity.

\begin{table}[t]
\centering

\begin{tabular}[t]{llllrrrrrr}
\toprule
$p$ & $n$ & $d$ & Model & \multicolumn{3}{c}{Average $\hat{d}$} & \multicolumn{3}{c}{Proportion $\hat{d} = d$}\\
\cmidrule(lr){5-7} \cmidrule(lr){8-10}
&&&& $\bZ$ & $\bX$ & $\widehat{\bX}$ & $\bZ$ & $\bX$ & $\widehat{\bX}$ \\
\midrule
40 & 200 & 1 & (i) & 1.00 & 1.00 & 1.00 & 1.00 & 1.00 & 1.00\\
 &  &  & (ii) & 1.00 & 1.00 & 1.00 & 1.00 & 1.00 & 1.00\\
 &  &  & (iii) & 1.00 & 1.00 & 1.00 & 1.00 & 1.00 & 1.00\\
 &  & 2 & (iv) & 1.76 & 1.96 & 1.71 & 0.75 & 0.93 & 0.70\\
 &  &  & (v) & 1.58 & 1.90 & 1.52 & 0.57 & 0.87 & 0.49\\
\addlinespace
40 & 500 & 1 & (i) & 1.00 & 1.00 & 1.00 & 1.00 & 1.00 & 1.00\\
 &  &  & (ii) & 1.00 & 1.00 & 1.00 & 1.00 & 1.00 & 1.00\\
 &  &  & (iii) & 1.00 & 1.00 & 1.00 & 1.00 & 1.00 & 1.00\\
 &  & 2 & (iv) & 1.75 & 1.96 & 1.72 & 0.75 & 0.96 & 0.72\\
 &  &  & (v) & 1.61 & 1.92 & 1.56 & 0.61 & 0.92 & 0.56\\
\addlinespace
500 & 200 & 1 & (i) & 1.00 & 1.00 & 1.00 & 1.00 & 1.00 & 1.00\\
 &  &  & (ii) & 1.58 & 1.11 & 1.01 & 0.53 & 0.89 & 0.97\\
 &  &  & (iii) & 1.55 & 1.12 & 1.02 & 0.56 & 0.90 & 0.97\\
 &  & 2 & (iv) & 2.08 & 1.77 & 1.61 & 0.37 & 0.54 & 0.57\\
 &  &  & (v) & 2.10 & 1.77 & 1.40 & 0.41 & 0.55 & 0.36\\
\addlinespace
1000 & 200 & 1 & (i) & 1.02 & 1.00 & 1.00 & 0.98 & 1.00 & 1.00\\
 &  &  & (ii) & 1.75 & 1.23 & 1.09 & 0.44 & 0.80 & 0.91\\
 &  &  & (iii) & 1.68 & 1.10 & 1.03 & 0.52 & 0.91 & 0.96\\
 &  & 2 & (iv) & 2.22 & 1.77 & 1.68 & 0.37 & 0.60 & 0.36\\
 &  &  & (v) & 2.10 & 1.87 & 1.50 & 0.41 & 0.45 & 0.36\\
\bottomrule
\end{tabular}
\caption{The mean estimated dimensions from three different lasso-\SIR estimators of $y$ on either the instrumental variables $\bZ$, the covariates subject to endogeneity $\bX$, or the first stage fitted values $\bXhat$. Standard errors are included in parentheses.}
\label{tab:selection_d}
\end{table}

Table \ref{tab:selection_d} shows that in settings with $p=40$ and $n\in\{200, 500\}$, all three methods performed well in selecting $d=1$ for the three single index models (i)--(iii), while the estimator as applied to the endogenous covariates $\bX$ surprisingly had the best performance for the double index models (iv)-(v). Turning to the high-dimensional settings with $ n=200$ and $p \in \{500, 1000\}$, all three methods still performed well for the linear model (i), but for the non-linear model (ii)-(iii), 
regressing $y$ on $\bZ$ and on $\bX$ tended to over-estimate $d$. On the other hand, for the double index models (iv)-(v), the lasso \SIR estimator on the instrumental variables $\bZ$ performed best while the other two approaches tended to underestimate the dimension. 

\section{Real data applications} \label{sec:applications}
We demonstrate our proposed method on two examples with endogeneity, one due to omitted variables and the other due to measurement error.

\subsection{Omitted variables: Mouse obesity data}
\label{subsection:data2}
We applied the proposed two-stage lasso \SIR estimator to perform simultaneous SDR and variable selection on a mouse obesity dataset from \citet{wang2006genetic}, and compared the results obtained to those from \citet{lin2015regularization} who analyzed the same data using their two-stage lasso estimator, as well as results based on applying the lasso \SIR estimator. Briefly, the study comprised $334$ mice fed a high-fat diet, and who were then genotyped using $1327$ single nucleotide polymorphisms (SNPs) with gene expressions profiled on microarrays that include probes for $23,388$ genes. We followed the same pre-processing steps as in \citet{lin2015regularization}, and obtained a complete dataset with $p=2825$ genes and $q=1250$ SNPs on $n=287$ mice. We treated the SNPs as instruments ($\bZ$), and focused on identifying important genes ($\bX$) related to the body weight ($y$). As mentioned in Section \ref{sec:intro}, the relationship between gene expression covariates and the outcome is likely confounded by omitted variables, experimental conditions, and environmental perturbations. As such, endogeneity due to omitted variable bias is expected to be a critical issue in this application. \citet{lin2015regularization} argued that genetic variants as represented by SNPs are valid instruments because the mice in this study come from the same genetic cross. 

\begin{figure}[htb]
\includegraphics[width = \textwidth]{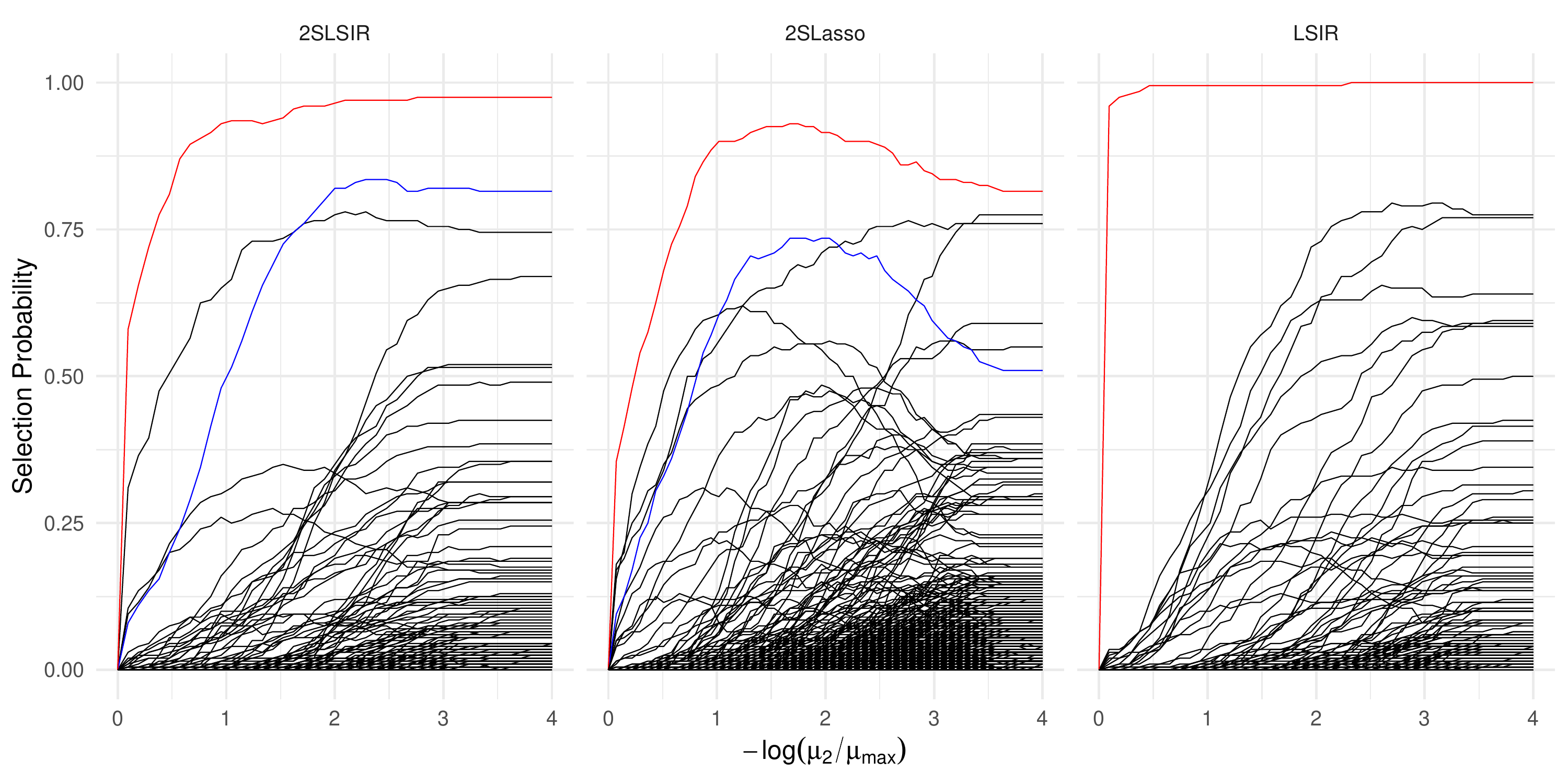}
\caption{Stability selection paths for three different methods applied to the mouse obesity data based on 100 random subsamples. With the maximum selection probability of at least 0.5, the gene selected by all the three methods is displayed in red, while the additional gene only selected by both two-stage methods is displayed in blue.} \label{fig:mouse-selection}
\end{figure}

To select the number of structural dimensions $d$, we applied the three approaches described in Section \ref{sec:selectiond}, and all selected $\hat{d} = 1$. Next, for each of the two-stage lasso, one-stage lasso \SIR, and two-stage lasso \SIR estimators, we applied stability selection to compute the empirical selection probability \citep{meinshausen2010stability} for each gene over 100 random subsamples of size $\lfloor n/2 \rfloor$ at each value of the tuning parameter. The upper bound for the per-family error rate was set to one, while the selection cutoff was set to $0.75$, meaning up to 75\% number of components can be included in the estimates in each subsample. The results in Figure \ref{fig:mouse-selection} present the selection path for each gene as a function of the second-stage tuning parameter $\mu_2$. With the maximum selection probability across all the tuning parameters set to $0.5$, the lasso \SIR estimator selected seven genes, the two-stage lasso selected nine genes, and the proposed two-stage lasso \SIR estimator selected six genes. Only one gene was chosen with a very high selection probability across all the three methods, while one additional gene was commonly selected by both the two-stage lasso \SIR and two-stage lasso estimators.

\subsection{Measurement error: NHANES data}
\label{subsec:data1}
The National Health and Nutrition Examination Survey (NHANES) aims to track the health and nutritional status of individuals in the United States  over time \citep{nhanesdata}. During the 2009-2010 survey period, participants were interviewed and asked to provide their demographic background, nutritional habits, and to undertake a series of health examinations. To assess the nutritional habits of participants, dietary data were collected using two 24-hour recall interviews, wherein the participants self-reported the consumed amounts of a set of food items during the 24 hours before each interview. Based on these recalls, daily aggregated consumption of water, food energy, and other nutrition components such as total fat and total sugar consumption were computed. In this section, we revisit the data application in \citet{nghiem2024likelihood}, seeking to understand the relationship between total cholesterol level ($y$) for $n = 3343$ women and their long-term intakes of $p=42$ nutrition components, particularly a linear combination/s of these components carrying the full regression information, and whether these combinations are sparse in nature.

Let $\bW_{ij}$ denote the vector of 42 nutritional covariates at the $j$th interview for $j=1, 2$. Dietary recall data are well-known to be subject to measurement errors, so we assume $\bW_{ij} = \bX_i + {E}_{ij}$, where $\bX_i$ represents the $i$th latent long-term nutritional intake, and ${E}_{ij}$ denotes an independent additive measurement error term following a multivariate Gaussian distribution with zero mean and covariance $\bSigma_u$. 
We used an instrumental variable approach to account for this measurement error. Specifically, following \citet[Section 6.5.1,][]{carroll2006measurement}, we treated the first replicate $\bW_{i1} = \bW_i$ as a surrogate for $\bX_i$ and the second replicate $\bW_{i2} = \bZ_i$ as an instrument. Note in this case, the number of instruments satisfies $q = p$. 
We then applied the linear instrumental variable model $\bX_i = \bGamma^\T \bZ_i + \bU_i$, where $\bGamma$ is allowed to be a $q\times p$ column-sparse matrix. It follows that we can write $\bW_i =  \bGamma^\T \bZ_i + \tilde{\bU}_i$ where $\tilde{\bU}_i = \bU_i + \bE_{i1}$. We then applied the proposed two-stage lasso \SIR estimator replacing $\mcX$ with
$\mathcal{W}$. All three methods for selecting the structural dimension described in Section \ref{sec:selectiond} selected $\hat{d} =1$; this was also consistent with the analysis in \citet{nghiem2024likelihood}. 

\begin{figure}[htb]
\centering
\includegraphics[width = 0.9\textwidth]{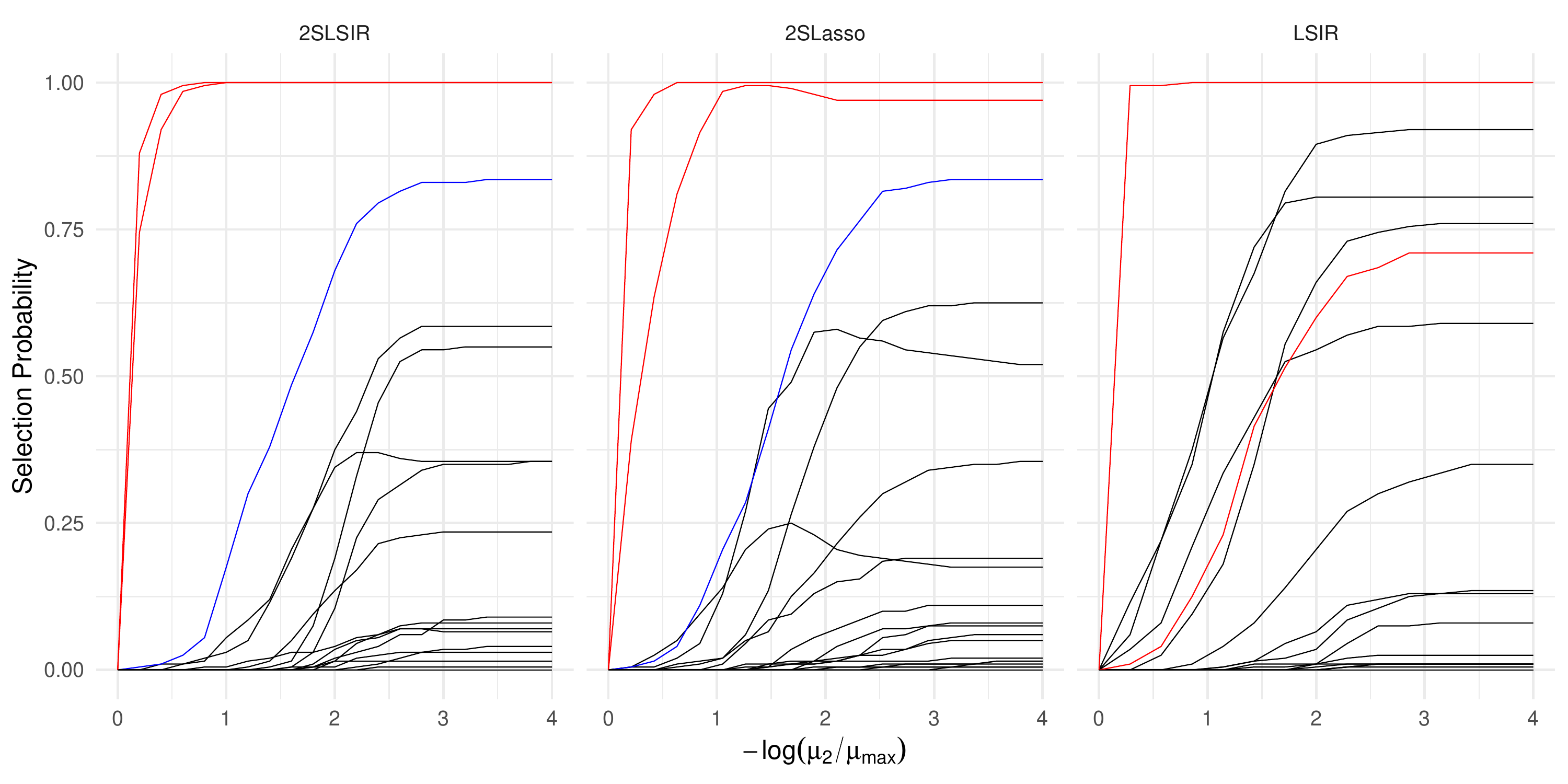}
\caption{Stability selection paths for three different methods applied to the NHANES data based on 100 random subsamples. With a maximum selection probability of at least 0.5, the components that are selected by all the three methods are displayed in red, while the additional component only selected by the two-stage methods is displayed in blue.}
\label{fig:nhanes}
\end{figure}

We first examined the estimated matrix coefficients $\widehat{\bGamma}$ from the first stage of the two-stage lasso \SIR estimation procedure. Figure 4 in Appendix D shows that the largest element in each column of $\widehat{\bGamma}$ is typically the diagonal element, so the most important instrument to model the consumption of any dietary component in the second interview is, unsurprisingly, the consumption of the same component in the first interview. Overall, each column of $\widehat{\bGamma}$ is relatively sparse, with the number of non-zero coefficients in each column ranging from 3 to 20.  
Next, 
we applied the same stability selection approach as in Section \ref{subsection:data2} to the proposed two-stage lasso \SIR estimator, and compared the resulting paths with the lasso \SIR and two stage lasso estimators.  
With the maximum selection probability across all the tuning parameters being at least 0.5, the lasso \SIR estimator selected five nutritional covariates, the two-stage lasso estimator selected five nutritional covariates, while the proposed two-stage lasso \SIR estimator selected six nutritional covariates. Figure \ref{fig:nhanes} presents the selection path as a function of the tuning parameter $\mu_2$ for each nutrition component. The two components commonly selected by all the three estimators were vitamin D and vitamin K, while both the two-stage lasso \SIR and two stage lasso estimators additionally select one more common component in folic acid. Vitamin D is the component with the largest estimated coefficients in both the two-stage lasso \SIR and two-stage lasso estimators, and was the second largest in the lasso \SIR estimator.

\section{Discussion} \label{sec:discussion}
In this paper, we propose a two-stage lasso \SIR estimator for simultaneous sufficient dimension reduction and variable selection in the presence of endogeneity. We establish estimation and selection consistency of the proposed estimator, allowing the dimensionality of both the covariates and instruments to grow exponentially with sample size. Numerical studies demonstrate the superior performance of the proposed estimator compared with existing methods that ignore non-linearity and/or endogeneity in the outcome model. 

Future research can extend the proposed method to accommodate a possible non-linear relationship between the covariates and the instruments in the first stage. Different sparsity structures on the coefficient matrix $\bGamma$, such as a low-rank matrix, can also be considered \citep{reinsel2022multivariate}. We can also examine endogeneity in other sufficient dimension reduction methods, such as inverse regression models \citep{bura2022sufficient}, and finally, we can adapt the methods to more complex settings involving endogeneity, such as SDR for clustered or longitudinal data \citep{hui2022sufficient,nghiem2024random}.

\bibliographystyle{apalike}

\bibliography{arxiv/main}

\begin{thebibliography}{}

\bibitem[Angrist and Krueger, 2001]{angrist2001instrument}
Angrist, J.~D. and Krueger, A.~B. (2001).
\newblock Instrumental variables and the search for identification: From supply
  and demand to natural experiments.
\newblock {\em Journal of Economic Perspectives}, 15(4):69--85.

\bibitem[Banerjee et~al., 2014]{estimation2014norm}
Banerjee, A., Chen, S., Fazayeli, F., and Sivakumar, V. (2014).
\newblock Estimation with norm regularization.
\newblock In Ghahramani, Z., Welling, M., Cortes, C., Lawrence, N., and
  Weinberger, K., editors, {\em Advances in Neural Information Processing
  Systems}, volume~27. Curran Associates, Inc.

\bibitem[Belloni et~al., 2012]{belloni2012sparse}
Belloni, A., Chen, D., Chernozhukov, V., and Hansen, C. (2012).
\newblock Sparse models and methods for optimal instruments with an application
  to eminent domain.
\newblock {\em Econometrica}, 80(6):2369--2429.

\bibitem[Bleazard et~al., 2015]{bleazard2015bias}
Bleazard, T., Lamb, J.~A., and Griffiths-Jones, S. (2015).
\newblock Bias in microrna functional enrichment analysis.
\newblock {\em Bioinformatics}, 31(10):1592--1598.

\bibitem[Bura et~al., 2022]{bura2022sufficient}
Bura, E., Forzani, L., Arancibia, R.~G., Llop, P., and Tomassi, D. (2022).
\newblock Sufficient reductions in regression with mixed predictors.
\newblock {\em Journal of Machine Learning Research}, 23(102):1--47.

\bibitem[Caner, 2009]{caner2009lasso}
Caner, M. (2009).
\newblock Lasso-type {GMM} estimator.
\newblock {\em Econometric Theory}, 25(1):270--290.

\bibitem[Caner et~al., 2018]{caner2018adaptive}
Caner, M., Han, X., and Lee, Y. (2018).
\newblock Adaptive elastic net {GMM} estimation with many invalid moment
  conditions: Simultaneous model and moment selection.
\newblock {\em Journal of Business \& Economic Statistics}, 36(1):24--46.

\bibitem[Card, 1999]{card1999causal}
Card, D. (1999).
\newblock The causal effect of education on earnings.
\newblock {\em Handbook of Labor Economics}, 3:1801--1863.

\bibitem[Carroll et~al., 2006]{carroll2006measurement}
Carroll, R.~J., Ruppert, D., and Stefanski, L.~A. (2006).
\newblock {\em Measurement error in nonlinear models}, volume 105.
\newblock CRC press.

\bibitem[{Centers for Disease Control and Prevention}, 2018]{nhanesdata}
{Centers for Disease Control and Prevention} (2018).
\newblock National health and nutrition examination survey data.
\newblock
  \url{https://wwwn.cdc.gov/nchs/nhanes/continuousnhanes/default.aspx?BeginYear=2009}.

\bibitem[Chernozhukov et~al., 2015]{chernozhukov2015post}
Chernozhukov, V., Hansen, C., and Spindler, M. (2015).
\newblock Post-selection and post-regularization inference in linear models
  with many controls and instruments.
\newblock {\em American Economic Review}, 105(5):486--490.

\bibitem[Cook and Forzani, 2008]{cook2008}
Cook, R.~D. and Forzani, L. (2008).
\newblock {Principal fitted components for dimension reduction in regression}.
\newblock {\em Statistical Science}, 23(4):485 -- 501.

\bibitem[Dai et~al., 2024]{dai2024new}
Dai, S., Wu, P., and Yu, Z. (2024).
\newblock New forest-based approaches for sufficient dimension reduction.
\newblock {\em Statistics and Computing}, 34(5):176.

\bibitem[Fan and Liao, 2014]{fan2014endogeneity}
Fan, J. and Liao, Y. (2014).
\newblock Endogeneity in high dimensions.
\newblock {\em Annals of Statistics}, 42(3):872.

\bibitem[Friedman et~al., 2010]{friedman2010regularization}
Friedman, J., Hastie, T., and Tibshirani, R. (2010).
\newblock {Regularization paths for generalized linear models via coordinate
  descent}.
\newblock {\em Journal of Statistical Software}, 33:1–22.

\bibitem[Gao et~al., 2020]{gao2020endogeneity}
Gao, J., Kim, N., and Saart, P.~W. (2020).
\newblock On endogeneity and shape invariance in extended partially linear
  single index models.
\newblock {\em Econometric Reviews}, 39(4):415--435.

\bibitem[Hansen et~al., 2008]{hansen2008estimation}
Hansen, C., Hausman, J., and Newey, W. (2008).
\newblock Estimation with many instrumental variables.
\newblock {\em Journal of Business \& Economic Statistics}, 26(4):398--422.

\bibitem[Horn and Johnson, 2012]{horn2012matrix}
Horn, R.~A. and Johnson, C.~R. (2012).
\newblock {\em Matrix analysis}.
\newblock Cambridge university press.

\bibitem[Hsing and Carroll, 1992]{hsing1992asymptotic}
Hsing, T. and Carroll, R.~J. (1992).
\newblock An asymptotic theory for sliced inverse regression.
\newblock {\em The Annals of Statistics}, pages 1040--1061.

\bibitem[Hu et~al., 2015]{hu2015identification}
Hu, Y., Shiu, J.-L., and Woutersen, T. (2015).
\newblock Identification and estimation of single-index models with measurement
  error and endogeneity.
\newblock {\em The Econometrics Journal}, 18(3):347--362.

\bibitem[Li, 2018]{li2018sufficient}
Li, B. (2018).
\newblock {\em {Sufficient dimension reduction: Methods and applications with
  R}}.
\newblock CRC Press, Florida.

\bibitem[Li et~al., 2017]{li2017feature}
Li, J., Cheng, K., Wang, S., Morstatter, F., Trevino, R.~P., Tang, J., and Liu,
  H. (2017).
\newblock Feature selection: A data perspective.
\newblock {\em ACM computing surveys (CSUR)}, 50(6):1--45.

\bibitem[Li, 1991]{li1991sliced}
Li, K.-C. (1991).
\newblock {Sliced inverse regression for dimension reduction}.
\newblock {\em Journal of the American Statistical Association}, 86:316–327.

\bibitem[Lin et~al., 2018]{lin2018consistency}
Lin, Q., Zhao, Z., and Liu, J.~S. (2018).
\newblock {On consistency and sparsity for sliced inverse regression in high
  dimensions}.
\newblock {\em The Annals of Statistics}, 46:580--610.

\bibitem[Lin et~al., 2019]{lin2019sparse}
Lin, Q., Zhao, Z., and Liu, J.~S. (2019).
\newblock Sparse sliced inverse regression via lasso.
\newblock {\em Journal of the American Statistical Association},
  114(528):1726--1739.

\bibitem[Lin et~al., 2015]{lin2015regularization}
Lin, W., Feng, R., and Li, H. (2015).
\newblock Regularization methods for high-dimensional instrumental variables
  regression with an application to genetical genomics.
\newblock {\em Journal of the American Statistical Association},
  110(509):270--288.

\bibitem[Ma and Zhu, 2013]{ma2013review}
Ma, Y. and Zhu, L. (2013).
\newblock {A review on dimension reduction}.
\newblock {\em International Statistical Review}, 81:134–150.

\bibitem[Meinshausen and B{\"u}hlmann, 2010]{meinshausen2010stability}
Meinshausen, N. and B{\"u}hlmann, P. (2010).
\newblock Stability selection.
\newblock {\em Journal of the Royal Statistical Society Series B: Statistical
  Methodology}, 72(4):417--473.

\bibitem[Nghiem and Hui, 2024]{nghiem2024random}
Nghiem, L.~H. and Hui, F. K.~C. (2024).
\newblock Random effects model-based sufficient dimension reduction for
  independent clustered data.
\newblock {\em Journal of Americal Statistical Association}, Accepted for
  publication.

\bibitem[Nghiem et~al., 2023a]{nghiem2024likelihood}
Nghiem, L.~H., Hui, F. K.~C., Muller, S., and Welsh, A.~H. (2023a).
\newblock Likelihood-based surrogate dimension reduction.
\newblock {\em Statistics and Computing}, 34(1):51.

\bibitem[Nghiem et~al., 2023b]{nghiem2023screening}
Nghiem, L.~H., Hui, F. K.~C., M{\"u}ller, S., and Welsh, A.~H. (2023b).
\newblock Screening methods for linear errors-in-variables models in high
  dimensions.
\newblock {\em Biometrics}, 79(2):926--939.

\bibitem[Nghiem et~al., 2023c]{nghiem2021sparse}
Nghiem, L.~H., Hui, F. K.~C., M{\"u}ller, S., and Welsh, A.~H. (2023c).
\newblock Sparse sliced inverse regression via cholesky matrix penalization.
\newblock {\em Statistica Sinica}, 33:1--33.

\bibitem[Ravikumar et~al., 2011]{ravikumar2011high}
Ravikumar, P., Wainwright, M.~J., Raskutti, G., and Yu, B. (2011).
\newblock High-dimensional covariance estimation by minimizing
  {$\ell_1$}-penalized log-determinant divergence.
\newblock {\em Electronic Journal of Statistics}, 5.

\bibitem[Sheng and Yin, 2016]{sheng2016sufficient}
Sheng, W. and Yin, X. (2016).
\newblock Sufficient dimension reduction via distance covariance.
\newblock {\em Journal of Computational and Graphical Statistics},
  25(1):91--104.

\bibitem[Tan et~al., 2018]{tan2018convex}
Tan, K.~M., Wang, Z., Zhang, T., Liu, H., and Cook, R.~D. (2018).
\newblock A convex formulation for high-dimensional sparse sliced inverse
  regression.
\newblock {\em Biometrika}, 105(4):769--782.

\bibitem[Wainwright, 2019]{wainwright2019high}
Wainwright, M.~J. (2019).
\newblock {\em High-dimensional statistics: A non-asymptotic viewpoint},
  volume~48.
\newblock Cambridge University Press.

\bibitem[Wang et~al., 2006]{wang2006genetic}
Wang, S., Yehya, N., Schadt, E.~E., Wang, H., Drake, T.~A., and Lusis, A.~J.
  (2006).
\newblock Genetic and genomic analysis of a fat mass trait with complex
  inheritance reveals marked sex specificity.
\newblock {\em PLoS Genetics}, 2(2):e15.

\bibitem[Windmeijer et~al., 2019]{windmeijer2019use}
Windmeijer, F., Farbmacher, H., Davies, N., and Davey~Smith, G. (2019).
\newblock On the use of the lasso for instrumental variables estimation with
  some invalid instruments.
\newblock {\em Journal of the American Statistical Association},
  114(527):1339--1350.

\bibitem[Xia et~al., 2002]{xia2002adaptive}
Xia, Y., Tong, H., Li, W., and Zhu, L.-X. (2002).
\newblock {An adaptive estimation of dimension reduction space}.
\newblock {\em Journal of the Royal Statistical Society Series B},
  64:363–410.

\bibitem[Xu et~al., 2022]{xu2022distributed}
Xu, K., Zhu, L., and Fan, J. (2022).
\newblock Distributed sufficient dimension reduction for heterogeneous massive
  data.
\newblock {\em Statistica Sinica}, 32:2455--2476.

\bibitem[Yang, 1977]{yang1977general}
Yang, S.-S. (1977).
\newblock General distribution theory of the concomitants of order statistics.
\newblock {\em The Annals of Statistics}, pages 996--1002.

\bibitem[Zhou and Zhu, 2016]{zhou2016principal}
Zhou, J. and Zhu, L. (2016).
\newblock Principal minimax support vector machine for sufficient dimension
  reduction with contaminated data.
\newblock {\em Computational Statistics \& Data Analysis}, 94:33--48.

\bibitem[Zhu et~al., 2006]{zhu2006sliced}
Zhu, L., Miao, B., and Peng, H. (2006).
\newblock On sliced inverse regression with high-dimensional covariates.
\newblock {\em Journal of the American Statistical Association},
  101(474):630--643.

\bibitem[Zou, 2006]{zou06}
Zou, H. (2006).
\newblock {The adaptive lasso and its oracle properties}.
\newblock {\em Journal of the American Statistical Association},
  101:1418–1429.

\end{thebibliography}

\appendix

\section{Performance of Lasso SIR under endogenity}

In this subsection, we consider a simple setting to demonstrate the performance of Lasso SIR when endogeneity is present. Suppose the $(p+1)-$random vector $(\bX, \varepsilon)^\top$ has the multivariate Gaussian distribution with zero mean, marginal covariance matrix $\bSigma_X = \bI_p$, and the marginal variance $\Var(\varepsilon) = 1$. In this case, it follows that
\begin{equation}
\begin{bmatrix}
\bX \\
\varepsilon\\
\end{bmatrix} \sim N_{p+1} \left(0, \begin{bmatrix}
\bI_p & \bSigma_{\bX, \varepsilon} \\
\bSigma_{\bX, \varepsilon}^\top & 1  
\end{bmatrix}
\right),
\label{eq:multivariateGaussian}
\end{equation}
where $\bSigma_{\bX, \varepsilon}$ is the correlation between $\bX$ and $\varepsilon$. Furthermore, assume that the outcome $y$ is generated from a linear model 
$
y = \bX \bbeta + \varepsilon.
$
Then we have
\[
\begin{bmatrix}
\bX \\
y\\
\end{bmatrix} \sim N_{p+1} \left(\boldsymbol{0}, \begin{bmatrix}
\bI_p & \bbeta + \bSigma_{\bX, \varepsilon} \\
\bbeta^\top + \bSigma_{\bX, \varepsilon}^\top & \bbeta^\top \bbeta + 2 \bSigma_{\bX, \varepsilon} \bbeta + 1  
\end{bmatrix}
\right),
\]
and, as a result, 
\[
E(\bX \mid y) = \dfrac{1}{ \bbeta^\top \bbeta + 2 \bSigma_{\bX, \varepsilon} \bbeta + 1}  (\bbeta + \bSigma_{\bX, \varepsilon}) y,
\]
so the eigenvector $\boldeta^*$ of $\bLambda^* = \Cov\left\{E(\bX \mid y)\right\}$ is the normalized eigenvector of $\bbeta + \bSigma_{\bX, \varepsilon}$. 

To demonstrate the performance of Lasso SIR, we conducted a small simulation study where we set $p=4$ and $\bbeta = (1, 1, 0, 0)^\top$, so the target of estimation for Lasso SIR is both the column space of $\bbeta$ as well as its support set $\mathcal{S} = \left\{1,2\right\}$. We generated $(\bX_i, \varepsilon_i)$ from the multivariate Gaussian distribution \eqref{eq:multivariateGaussian}, with  three scenarios for the correlation vector $\bSigma_{\bX, \varepsilon}$, including
(I) $\bSigma_{\bX, \varepsilon} = (0.5, 0.5, 0, 0)^\top$, (II) $\bSigma_{\bX, \varepsilon} = (0.5, -0.5, 0, 0)^\top$, and (III) 
$\bSigma_{\bX, \varepsilon} = (-0.5, -0.5, 0.5, 0.5)^\top$. 
We then simulated the outcome from either the linear model $y = \bX \bbeta + \varepsilon$ or the non-linear model $y = \sin(\bX\bbeta + \varepsilon)$. We varied $n \in \{100, 500, 1000\}$ and for each simulation configuration, we generated $1000$ samples.   

For each generated sample, we computed the Lasso SIR estimator $\hat{\bbeta}^*$ with the tuning parameter $\mu$ selected by 10-fold cross-validation, as implemented by the \texttt{LassoSIR} package. We reported the estimation error in estimating the column space spanned by $\bbeta$, as measured by $\norm{\mathcal{P}(\bbeta) - \mathcal{P}(\hat\bbeta)}_F$. For variable selection performance, we computed the area under the receiver operating curve (AUC); an AUC of one indicates a clear separation between the estimates of non-important variables and the those of important variables in $\hat{\bbeta}^*$.

Table \ref{tab:LassoSIR} shows that the performance of the Lasso SIR estimator aligns with our theoretical results. In the first case, the correlation vector $\bSigma_{\bX, \varepsilon}$ is proportional to the true $\bbeta$, so the Lasso SIR maintains both estimation and variable selection consistency, as evidenced by a decreasing estimation error when $n$ increases and an AUC of 1.  In the second case, the Lasso SIR estimator gives a consistent estimate of the space spanned by $\boldeta^* = (1.5, 0.5, 0, 0)/\sqrt{2.25}$, which has a different direction but the same support compared to the true $\bbeta$. Hence, the Lasso SIR is not estimation consistent, but still has an AUC of 1.  
In the last case, the Lasso SIR estimator gives a consistent estimate for the space spanned by $\boldeta^* = (0.5, 0.5, 0.5, 0.5)^\top$, which has both a different direction and a different support from those of the true $\bbeta$. As a result, the Lasso SIR estimator is not selection or estimation consistent.

\begin{table}[th]
\centering
\caption{Performance of the Lasso SIR estimator under endogeneity, as measured by the estimation error $\norm{\mathcal{P}(\bbeta) - \mathcal{P}(\hat\bbeta^*)}_F$ and variable selection performance (AUC).}
\label{tab:LassoSIR}
\begin{tabular}[t]{rrrrrr}
\toprule
$\bSigma_{\bX, \varepsilon}$ & $n$ & \multicolumn{2}{c}{Linear link} & \multicolumn{2}{c}{Sine link}  \\
\addlinespace[.5pt]
& & Error & AUC & Error & AUC\\
\midrule
(I) & 100 & 0.095 & 1.000 & 0.876 & 0.853\\
& 500 & 0.039 & 1.000 & 0.402 & 0.992\\
 & 1000 & 0.028 & 1.000 & 0.258 & 1.000\\
\addlinespace
(II) & 100 & 0.659 & 1.000 & 0.768 & 0.907\\
 & 500 & 0.643 & 1.000 & 0.674 & 0.995\\
& 1000 & 0.639 & 1.000 & 0.661 & 1.000\\
\addlinespace
(III) & 100 & 1.001 & 0.758 & 1.003 & 0.739\\
& 500 & 1.000 & 0.744 & 0.999 & 0.756\\
& 1000 & 1.000 & 0.728 & 0.999 & 0.747\\
\bottomrule
\end{tabular}
\end{table}

\section{Technical Proofs}
\subsection{Proof of Proposition 1}
Without loss of generality, assume the expectation of $\text{E}(\bX, \varepsilon) = 0$, the marginal covariance matrix $\bSigma_X = \bI_p$, and the marginal variance $\Var(\varepsilon) = 1$. In this case, it follows that 
$$
\begin{bmatrix}
\bX \\
\varepsilon\\
\end{bmatrix} \sim N_{p+1} \left(\boldsymbol{0}, \begin{bmatrix}
\bI_p & \bSigma_{\bX, \varepsilon} \\
\bSigma_{\bX, \varepsilon}^\top & 1  
\end{bmatrix}
\right).
$$
By the law of iterated expectation, we have
$$
E(\bX \mid y) = E\left\{ E(\bX \mid \bB^\top \bX, \varepsilon, y)\mid y \right\} = E\left\{E(\bX \mid \bB^\top \bX, \varepsilon) \mid y \right\}.
$$
Next, by the property of the multivariate Gaussian distribution we obtain
$$
\small
\begin{aligned}
& E(\bX \mid \bB^\top \bX, \varepsilon)  = \begin{bmatrix}
\bB & \bSigma_{\bX, \varepsilon} 
\end{bmatrix} \begin{bmatrix}
\bB^\top \bB & \bB^\top  \bSigma_{\bX, \varepsilon} \\
 \bSigma_{\bX, \varepsilon}^\top \bB & 1 
\end{bmatrix}^{-1} \begin{bmatrix}
\bB^\top \bX \\
\varepsilon
\end{bmatrix} 
\\ & = \begin{bmatrix}
\bB & \bSigma_{\bX, \varepsilon} 
\end{bmatrix} \begin{bmatrix}
(\bB^\top \bB - \bB^\top \bSigma_{\bX, \varepsilon}\bSigma_{\bX, \varepsilon}^\top \bB)^{-1} &  -(\bB^\top \bB - \bB^\top \bSigma_{\bX, \varepsilon}\bSigma_{\bX, \varepsilon}^\top \bB)^{-1}\bB^\top  \bSigma_{\bX, \varepsilon} \\
 -\bSigma_{\bX, \varepsilon}^\top \bB (\bB^\top \bB - \bB^\top \bSigma_{\bX, \varepsilon}\bSigma_{\bX, \varepsilon}^\top \bB)^{-1}& 1 + \bSigma_{\bX, \varepsilon}^\top \bB (\bB^\top \bB - \bB^\top \bSigma_{\bX, \varepsilon}\bSigma_{\bX, \varepsilon}^\top \bB)\bB^\top  \bSigma_{\bX, \varepsilon}
\end{bmatrix}\begin{bmatrix}
\bB^\top \bX \\
\varepsilon
\end{bmatrix}
\end{aligned}
$$
which is in the subspace spanned by $\bB$ if and only if $\bSigma_{\bX, \varepsilon}$ can be written in the form $\bSigma_{\bX, \varepsilon} = \bB \bA$ for a $d\times d$ matrix $\bA$, i.e $\bSigma_{\bX, \varepsilon} \in \text{col}(\bB)$.

\subsection{Proof of Proposition 2}

Let $\tilde{\boldeta}_k^* = \boldeta_k^*\boldeta_k^{*\top} \hat{\boldeta}^*_k$, so each element of $\tilde{\boldeta}_k^*$ always has the same sign as that of $\hat{\boldeta}_k^*$. Let $\tilde\bbeta_k^* = \bSigma_\bX^{-1} \tilde{\boldeta}_k^* = \bbeta_k^*\boldeta_k^{*\top} \hat{\boldeta}^*_k$. Since $p$ is fixed, the cross product ${\boldeta}_k^{*\top}\hat{\boldeta}_{k}^{*} \neq 0$ with probability one for all $k=1,\ldots, d$. As a result, we have $\mathcal{P}(\widetilde{\bB}^*) = \mathcal{P}({\bB}^*)$ with probability one.  

From the definition of the Lasso-SIR estimator and writing $\hat\bbeta_k^* = \tilde{\bbeta}_k^* + \bm{\delta}_k^*$, we have
\[
\frac{1}{n} \norm{\tilde{\bm{y}}^*_k - \mcX \tilde{\bbeta}^*_k - \mcX \bdelta_k^*}^2 + \mu_k \|\tilde{\bbeta}^*_{k} + \bdelta_{k}^*\|_1 \leq \frac{1}{n} \norm{\tilde{\bm{y}}^*_k - \mcX {\bbeta}^*_k}^2 + \mu_k \|\tilde{\bbeta}^*_{k}\|_1 .
\]
Equivalently, we have 
\[
\bdelta_k^{*^\top}\left(\frac{\mcX^\top \mcX}{n} \right) \bdelta_k^*  - 2 \bdelta_k^{*^\top} \left(\dfrac{\mcX^\top\tilde{\bm{y}}^*_k}{n} - \dfrac{\mcX^\top \mcX}{n}\tilde{\bbeta}_k^*   \right) \leq \mu_k \left(\|\tilde{\bbeta}^*_{k}\|_1 - \|\tilde{\bbeta}^*_{k} + \bdelta_{k}^*\|_1 \right).
\]
By construction, we have $n^{-1} \mcX^\top \tilde{\by}_k^* = \hat{\boldeta}_k^*$, the eigenvector associated with the $k$th largest eigenvalue of $\hat{\bLambda}^*$. By the law of large numbers, we have $\hat{\bLambda}^* \xrightarrow{p} \bLambda^*$, so $\norm{\hat\boldeta_k^* - \tilde{\boldeta}_k^*} = o_p(1)$. Furthermore, as $n \to \infty$, we have $n^{-1}{\mcX^\top \mcX} \xrightarrow{p} \bSigma_{\bX}$
so $n^{-1} \mcX^\top \mcX \tilde{\bbeta}_k^* \xrightarrow{p} \tilde{\boldeta}_k^*$. 
Therefore, as $n \to \infty$. the vector $\bdelta_k^*$ satisfies
\[
\bdelta_k^{*^\top} \bSigma_{\bX} \bdelta_k^{*} \leq \mu_k \left(\|\tilde{\bbeta}^*_{k}\|_1 - \|\tilde{\bbeta}^*_{k} + \bdelta_{k}^*\|_1 \right) + 2\bdelta_k^*\left(\hat\boldeta_k^* - \tilde{\boldeta}_k^* \right).
\]
Since $\mu_k = o(1)$ and $\norm{\hat\boldeta_k^* - \tilde{\boldeta}_k^*} = o_p(1)$, we have $\norm{\bdelta_k^*} = o_p(1)$, so $\mathcal{P}(\bB_k^*) \xrightarrow{p} \mathcal{P}(\widetilde{\bB}^*) = \mathcal{P}({\bB}^*)$ as $n \to \infty$.

\subsection{Proof of Lemma 1}
The instrument $\bZ$ is independent of both $\varepsilon$ and $\bU$, so the conditional expectation  $\widehat{\bX} = E(\bX \mid \bZ) = \bGamma \bZ$ is also independent of both $\varepsilon$ and $\bU$. By the properties of conditional expectation, we have
\begin{align*}
E(\bXhat \mid y) &  = E\left[E \left(\bXhat \mid \bB^\top \bX, y \right) \mid y \right]\\
& = E\left[E \left(\bXhat \mid \bB^\top \bX, \varepsilon \right) \mid y \right] ~ \quad  \text{(since $y$ is a deterministic function of $\bX^\top\bB$ and $\varepsilon$}) \\
& = E \left[E\left(\bXhat \mid \bB^\top\bX\right) \mid y \right] ~ \quad  ~~~ \text{(since $\varepsilon$ and $\bXhat$ are independent)} \\
& \stackrel{(i)}{=} \Cov(\bXhat, \bX) \bB\left[ \Var(\bB^\top \bX) \right]^{-1}  \bB^\top E(\bX \mid y),
\end{align*}
where step $(i)$ follows from Condition 3.1 in the main text and Lemma 1.01 of \citet{li2018sufficient}. The last equation shows that the conditional expectation of $E(\bXhat \mid Y)$ is contained in the space spanned by $\bSigma_{\bXhat} \bB$, and the required result follows.

\subsection{Properties of $\hat{\bLambda}$}
\label{section:hatbLambda}


The developments below closely follow the work of \citet{tan2018convex}. Note 
$$
\bLambda = \Cov\left\{E\left(\widehat{\bX} \mid y\right) \right\} = \bGamma^\top \Cov\left\{E\left({\bZ} \mid y\right)\right\} \bGamma = \bGamma^\top \left(\bSigma_\bZ - {\bXi} \right) \bGamma,
$$ 
where $\bXi = E\left\{\Cov\left({\bZ} \mid y\right)\right\}$. Similarly, we can write $\widehat{\bLambda} = \widehat{\bGamma}^\top (\widehat{\bSigma}_{\bZ} - \widehat{\bXi}) \widehat{\bGamma}$, with 
\begin{equation}
\widehat{\bXi} = \dfrac{1}{H(c-1)} \mcZ^\top (\bI_H \otimes \bP_c) \mcZ, 
\end{equation}
with $\bP_c = \bI_c - \be_c\be_c^\top$, $\mathbf{e}_c = (1, \ldots, 1)^\top$ being the  $c\times 1$ vector of ones, and $\bI_c$ denoting the $c \times c$ identity matrix. Define $\bQ_c=\bP_c-\left(1-c^{-1}\right) \bI_c=c^{-1}\left(\bI_c-\mathbf{e}_c \mathbf{e}_c^T\right)$, and let $\bm{\Omega} = \bSigma_{\bZ} - \bXi$, and $\widehat{\bm{\Omega}} = \widehat{\bSigma}_{\bZ} - \widehat{\bXi}$. 

We now bound the elementwise maximum norm $\norm{\widehat{\bLambda} - \bLambda}_{\max}$. By the triangle inequality, for any $j, k = 1,\ldots, q$, we have 
\begin{equation}
\begin{aligned}
& \vert \bgamma_j^\top \bOmega \bgamma_k - \hat{\bgamma}_j^\top \widehat{\bOmega}  \hat{\bgamma}_k \vert  \\ & =   \vert (\bgamma_j^\top \bOmega \bgamma_k  - \hat{\bgamma}_j^\top \bOmega \bgamma_k)  + (\hat{\bgamma}_j^\top \bOmega \bgamma_k - \hat{\bgamma}_j^\top \widehat{\bOmega} \bgamma_k)  + (\hat{\bgamma}_j^\top \widehat{\bOmega} \bgamma_k  - \hat{\bgamma}_j^\top \widehat{\bOmega} \hat{\bgamma}_k) \vert \\& \leq \vert \bgamma_j^\top {\bOmega} \bgamma_k  - \hat{\bgamma}_j^\top {\bOmega} \bgamma_k \vert + \vert \hat{\bgamma}_j^\top {\bOmega} \bgamma_k - \hat{\bgamma}_j^\top \widehat{\bOmega} \bgamma_k \vert + \vert \hat{\bgamma}_j^\top \widehat{\bOmega} \bgamma_k  - \hat{\bgamma}_j^\top \widehat{\bOmega} \hat{\bgamma}_k \vert \\
& \leq \Vert \hat{\bgamma}_j-\bgamma_j \Vert_1 \Vert {\bOmega}\bgamma_k \Vert_\infty + \Vert \hat\bgamma_j \Vert_1 \Vert \bOmega - \widehat{\bOmega} \Vert_{\max} \Vert \bgamma_k \Vert_1 +  \Vert \widehat{\bOmega} \hat\bgamma_j  \Vert_\infty \Vert  \hat{\bgamma}_k-\bgamma_k \Vert_1  \\&   
\leq \Vert \hat{\bgamma}_j-\bgamma_j \Vert_1 \norm{\bOmega}_{\max} \norm{\bgamma_k}_1 + \Vert \hat\bgamma_j \Vert_1 \Vert \bOmega - \widehat{\bOmega} \Vert_{\max} \Vert \bgamma_k \Vert_1 +  \norm{\widehat{\bOmega}}_{\max} \norm{\hat\bgamma_j}_1  \Vert  \hat{\bgamma}_k-\bgamma_k \Vert_1 
\\ & = T_1 + T_2 + T_3.
\end{aligned}
\label{eq:crossprod}
\end{equation}

From Theorem 1 in the main text we have $\Vert \hat{\bgamma}_j-\bgamma_j \Vert_1 = C \sigma_{\max}r \sqrt{\log(pq)/n}$ with probability at least $1-\exp(-C^\prime  \log pq)$ for any $j \in 1, \ldots, q$. Furthermore by Condition 4.2 in the main text, we have $\norm{\bgamma_k}_1 \leq L$ and $\Vert \bOmega \Vert_{\max} \leq C$. Therefore, obtain
\begin{equation}
T_1 \leq CL\sigma_{\max}r\sqrt{\frac{ \log pq}{n}},
\label{eq:T1section1}
\end{equation}
For the second term $T_2$, by the triangle inequality we have that
$$
\Vert \hat\bgamma_j \Vert_1 = \Vert \hat\bgamma_j - \bgamma_j + \bgamma_j \Vert_1 \leq \Vert \hat\bgamma_j - \bgamma_j \Vert_1 + \Vert \bgamma_j \Vert_1 \leq C{\sigma_{\max}}r  \sqrt{\dfrac{\log pq}{n}} + L \leq C^{\prime}L,
$$
with probability at least $1-\exp(-C^{\prime\prime} \log pq)$, assuming ${\sigma_{\max}}r \sqrt{\log q/{n}}
= o(1)$. Next, Lemma \ref{lemma:boundingnormmax} provides an upper bound on $\Vert \bOmega - \widehat{\bOmega} \Vert_{\max}$ as follows.
\begin{lemma}
Under the Conditions 4.1-4.5 in the main text, we have $\Vert \bOmega - \widehat{\bOmega} \Vert_{\max} \leq C \sqrt{\log (q)/n}$ with probability at least $1-\exp(-C^\prime \log q)$.
\label{lemma:boundingnormmax}
\end{lemma}
The proof for Lemma \ref{lemma:boundingnormmax}and other technical results in this section are deferred to section \ref{sec:additionalproofs}.
Combining Lemma 1 and union bound then, we obtain
\begin{equation}
T_2 \leq CL^2\sqrt{\dfrac{\log q}{n}}.
\label{eq:T2section1}
\end{equation}
with a probability at least $1-\exp(-C^\prime \log q) - \exp(-C^{\prime\prime} \log pq)$. 

Finally for the term $T_3$, by the triangle inequality we also have
$$\norm{\widehat{\bOmega}}_{\max} \leq \norm{\widehat{\bOmega} - \bOmega}_{\max} + \norm{{\bOmega}}_{\max} \leq C,
$$
with a probability at least $1-\exp(-C^{\prime} \log q)$, assuming $(\log q)/n = o(1)$. Hence, with a probability of at least $1-\exp(-C^\prime \log q) - \exp(-C^{\prime\prime} \log pq)$, we have
\begin{equation}
T_3 \leq CL\sigma_{\max} r \sqrt{\dfrac{\log pq}{n}}.
\label{eq:T3section1}
\end{equation}

Substituting \eqref{eq:T1section1}-\eqref{eq:T3section1} into \eqref{eq:crossprod} and taking the union bound, we finally obtain
\begin{equation}
\norm{\widehat{\bLambda} - \bLambda}_{\max} = \max_{j,k}   \vert \bgamma_j^\top \bOmega \bgamma_k - \hat{\bgamma}_j^\top \widehat{\bOmega}  \hat{\bgamma}_k \vert \leq CL\left\{L\sqrt{\dfrac{ \log q}{n}} +   \sigma_{\max} r \sqrt{\dfrac{\log pq}{n}} \right\}, 
\end{equation}
with a probability at least $1-\exp(-C^\prime \log q) - \exp(-C^{\prime\prime}\log pq)$.

\bigskip
\underline{Eigenvalue and eigenvector analyses for $\hat{\bLambda}$}

Recall that at the population level ${\bLambda} = {\bGamma}^\top {\bOmega} \bGamma$, where $\bOmega$ has rank $d$. Let $\boldeta_j(\bOmega)$ be the eigenvector associated with the $j$th largest eigenvalue $\lambda_j(\bOmega)$ of $\bOmega$ for $j=1,\ldots, d$, and we also write $\lambda(\bOmega) = \lambda_1(\bOmega)$ and $\boldeta(\bOmega) = \boldeta_1(\bOmega)$ (i.e we suppress the subscript 1 when we refer to the largest eigenvalue). When $d=1$, the (only) non-zero eigenvalue of $\bLambda$ is $\lambda(\bLambda)  = \lambda(\bOmega) \Vert \bGamma^\top \boldeta(\bOmega)\Vert_2$, but if $d >1$, there are generally no closed form for the eigenvalues of $\bLambda$. Nevertheless, because $\bOmega$ has rank $d \leq \min(p, q)$, applying the Weyl's inequality yields
\begin{equation}
\nu_{\min}^2 (\bGamma) \lambda_j(\bOmega) \leq \lambda_j(\bLambda) \leq \nu_{\max}^2 (\bGamma) \lambda_j(\bOmega), 
\label{eq:weylinequality}
\end{equation}
for $j=1,\ldots, d$, where $\nu_{\min}(\bGamma)$ and $\nu_{\max}(\bGamma)$ denote the minimum singular value and maximum singular value for $\bGamma$. Similar inequalities hold for $\widehat{\bLambda}$, respectively. Furthermore, since conditions 4.4--4.6 are imposed, the results from \citet{lin2019sparse} are applied to various quantities related to $\widehat{\bOmega}$, which we state in Proposition \ref{prop:yonZ} below. 

\begin{proposition}
\label{prop:yonZ}
   If $n\lambda(\bOmega)=q^\alpha$ for some $\alpha>1 / 2$, there exist positive constants $C_1$ and $C_2$ such that, with a probability of at least $1-\exp(C^{\prime} \log(q))$
   
i) for $j=1, \ldots ., d$, one has
$$
\Vert P_{\bOmega}{\boldeta}_j(\widehat{\bOmega})\Vert_2 \geq C_1 \sqrt{\frac{\lambda(\bOmega)}{{\lambda}_j(\widehat{\bOmega})}}~~,
$$

ii) for $j=d+1, \ldots, H$, one has
$$
\Vert P_{\bOmega}{\boldeta}_j(\widehat{\bOmega})\Vert_2 \leq C_2 \frac{\sqrt{q \log (q)}}{n \lambda(\bOmega)} \sqrt{\frac{\lambda(\bOmega)}{{\lambda_j}(\widehat{\bOmega})}}.
$$
\end{proposition}
With these results, combining with the results from Theorem 1, we then have for $j=1,\ldots, d$,
\begin{equation}
\dfrac{\lambda(\bLambda)}{\lambda_j(\bhatLambda)} \leq \dfrac{\lambda(\bOmega)}{\lambda_j(\widehat{\bOmega})} \times  \dfrac{\nu_{\max}(\bGamma)}{\nu_{\min}(\widehat{\bGamma})} \leq C, 
\end{equation}
with probability of at least $1-\exp(-C^\prime \log q) - \exp(-C^{\prime\prime}\log pq)$.

\subsection{Proof of Lemma 3}
By the optimality of $\hat{\bbeta}$, we  obtain
\begin{equation}
\dfrac{1}{2n} \Vert \tilde{\bby}  - \hat{\mcX}\hat{\bbeta} \Vert _2^2 + \mu_2 \Vert \hat\bbeta \Vert_1 \leq \dfrac{1}{2n} \Vert \tilde{\bby}  - \hat{\mcX}{\tilde\bbeta} \Vert _2^2 + \mu_2 \Vert {\tilde\bbeta} \Vert_1.  
\label{eq:optimality}
\end{equation}
Letting $\bdelta = \hat\bbeta - \tilde\bbeta$ and substituting into \eqref{eq:optimality}, we obtain
$$
\dfrac{1}{2n} \Vert \tilde{\bby}  - \hat{\mcX}\tilde{\bbeta} -\hat{\mcX}\bdelta \Vert _2^2 + \mu_2 \Vert \bdelta +  \tilde\bbeta \Vert_1 \leq \dfrac{1}{2n} \Vert \tilde{\bby}  - \hat{\mcX}{\tilde\bbeta} \Vert _2^2 + \mu_2 \Vert {\tilde\bbeta} \Vert_1.  
$$
Expanding the square, cancelling the same terms on both sides, and rearranging terms, we have
\begin{equation}
\dfrac{1}{2n} \Vert \hat{\mcX}\bdelta \Vert_2^2 - \dfrac{1}{n} \bdelta^\top \hat{\mcX}^\top(\tilde{\bby}  - \hat{\mcX}\tilde\bbeta) \leq \mu_2 \left(\Vert \tilde\bbeta\Vert_1 - \Vert \tilde \bbeta + \bdelta  \Vert_1 \right). 
\label{eq:4}
\end{equation}
By definition, we have $\hat{\bm\eta}  = n^{-1}\hat{\mcX}^\top \tilde\bby $, and so \eqref{eq:4} is equivalent to
$$
\dfrac{1}{2n} \Vert \hat{\mcX}\bdelta \Vert_2^2 \leq  \bdelta^\top \left(\hat\boldeta  - \dfrac{1}{n} \hat{\mcX}^\top \hat{\mcX} \tilde\bbeta\right) + \mu_2 \left(\Vert \tilde\bbeta\Vert_1 - \Vert \tilde \bbeta + \bdelta  \Vert_1 \right). 
$$
Applying H\"{o}lder's inequality and the condition on $\mu_2$ from the Lemma, we obtain
$$
\begin{aligned}
 \bdelta^\top \left(\hat\boldeta  - \dfrac{1}{n} \hat{\mcX}^\top \hat{\mcX} \tilde\bbeta\right) & \leq \Vert \bdelta \Vert_1 \left\Vert \hat\boldeta  - \dfrac{1}{n} \hat{\mcX}^\top \hat{\mcX} \tilde\bbeta \right\Vert_\infty \leq \dfrac{1}{2} \mu_2 \Vert \bdelta \Vert_1. 
\end{aligned}
$$
It follows that
\begin{equation}
\frac{1}{2} \Vert \bdelta \Vert_1 + \left(\Vert \tilde\bbeta\Vert_1 - \Vert \tilde \bbeta + \bdelta  \Vert_1 \right) \geq 0.
\label{eq:delta}
\end{equation}
Furthermore, applying the (reverse) triangle inequality gives
\begin{align*}
\norm{\tilde{ {\bbeta}}}_1-\norm{\tilde{\bbeta} + \bdelta}_1  = \norm{\tilde{ {\bbeta}}_S}_1 - \norm{\tilde{ {\bbeta}}_S+ {\bdelta}_S}_1 - \norm{ \bdelta_{S^c}}_1 \leq \norm{ {\bdelta}_S}_1 -\norm{ {\bdelta}_{S^c}}_1.
\end{align*} 
Since $\norm{ {\bdelta}}_1 = \norm{ {\bdelta}_S}_1+\norm{ {\bdelta}_{S^c}}_1$, then equation \eqref{eq:delta} further leads to
\begin{align*}
0 & \leq \dfrac{1}{2}\norm{ {\bdelta}}_1+\norm{\tilde{ {\bbeta}}}_1-\norm{\tilde{\bbeta} + \bdelta}_1 \\ & \stackrel{(iv)}{\leq} \dfrac{1}{2}{\norm{ {\bdelta}_S}_1+\dfrac{1}{2}\norm{ {\bdelta}_{S^c}}_1} +\norm{ {\bdelta}_S}_1 -\norm{ {\bdelta}_{S^c}}_1  = \dfrac{3}{2}\norm{ {\bdelta}_{S}}_1 - \dfrac{1}{2} \norm{ {\bdelta}_{S^c}}_1  \\& \stackrel{(v)}{\leq} \dfrac{3}{2}\norm{ {\bdelta}_{S}}_1. 
\end{align*}
It follows from step (iv) that $\norm{ {\bdelta}_{S^c}}_1 \leq 3 \norm{ {\bdelta}_{S}}_1$, or $ \bdelta \in \mathcal{C}(S,3)$. 

Next, we apply the following result which establishes the sample covariance matrix $n^{-1} \widehat{\mcX}^\top\widehat{\mcX}$ satisfies the RE condition with a parameter $\kappa_2$ with high probability.

\begin{proposition}
Assume Conditions 4.1-4.2, 4.5 and the requirements of Theorem 1 are satisfied. Then there exists a constant $\kappa_2 > 0$ such that $\norm{\widehat{\bSigma}_{\bXhat}^{1/2} \bdelta}_2^2 \geq \kappa_2 \norm{\bdelta}_2^2$, for any $\delta \in \mathcal{C}(S, 3)$, with probability at least $1- \exp(-C^\prime \log q) - \exp(-C^{\prime\prime} \log pq)$, for constants $C^\prime$ and $C^{\prime\prime}$. 
\label{prop:Re2}
\end{proposition}

Conditioned on this RE condition, step (v) implies
\[
\begin{split}
\dfrac{1}{2} \kappa_2 \norm{ \bdelta}_2^2 {\leq} \dfrac{1}{2n}\norm{\hat{\mcX} {\bdelta}}_2^2 & {\leq} \dfrac{1}{2}\mu_2\norm{ {\bdelta}}_1  + \mu_2 \left(\norm{\tilde{ {\bbeta}}}_1-\norm{\hat{ {\bbeta}}}_1 \right)\\& {\leq} \dfrac{3}{2} \mu_2\norm{ \bdelta_S}_1 {\leq} \dfrac{3}{2}\mu_2\sqrt{s}\norm{ \bdelta_S}_2 \leq \dfrac{3}{2}\mu_2\sqrt{s}\norm{ \bdelta}_2.
\end{split}
\]
As a result, we obtain 
$$
\norm{\hat{ \bbeta}-\tilde{ \bbeta}}_2 = \norm{ \bdelta}_2 \leq \dfrac{3}{\kappa_2}\mu_2 \sqrt{s},
$$ 
as required.

\subsection{Proof of Theorem 2}

Recall that $\tilde{\bbeta} = \xi^{-1}\bbeta (\hat{\boldeta}^\top \boldeta)$ with $\xi = \norm{\Sigma_{\bXhat}\bbeta}_2 > 0$. Since the quantity $\hat{\boldeta}^\top \boldeta$ is non-zero with a probability one, so $P(\bbeta) = P(\hat{\bbeta})$ with the same probability. By the Cauchy-Schwartz inequality we have
$$
\left\|P(\hat\bbeta) - P({\bbeta})\right\|_F=\left\|P(\hat\bbeta) - P({\tilde\bbeta})\right\|_F \leq 4 \frac{\|\hat{\bbeta}-\tilde{\bbeta}\|_2} 
{\|\tilde{\bbeta}\|_2} = C\|\hat{\bbeta}-\tilde{\bbeta}\|_2.
$$
Thus the required result will follow from Lemma 3.2, if the choice of $\mu_2$ as given in theorem satisfies $\mu_2 \geq 2 \Vert \hat\boldeta - n^{-1} \hat{\mcX}^\top \hat{\mcX} \tilde\bbeta \Vert_\infty$ with probability tending to one. It remains then to find an upper bound for $\Vert \hat\boldeta - n^{-1} \hat{\mcX}^\top \hat{\mcX} \tilde\bbeta \Vert_\infty$. 

By the triangle inequality, we have
\begin{align}
\left\Vert \hat\boldeta - \dfrac{1}{n} \hat{\mcX}^\top \hat{\mcX} \tilde\bbeta \right\Vert_\infty & = \left\Vert \hat\boldeta - \bSigma_{\bXhat}\tilde\bbeta + \bSigma_{\bXhat}\tilde\bbeta - \dfrac{1}{n} \hat{\mcX}^\top \hat{\mcX} 
\tilde\bbeta \right\Vert_\infty \nonumber \\
& \leq \underbrace{\left\Vert \hat\boldeta - \bSigma_{\bXhat}\tilde\bbeta \right\Vert_\infty}_{\text{I}} + \underbrace{\left\Vert \left(\bSigma_{\bXhat} - \widehat{\bSigma}_{\bXhat}\right)
\tilde\bbeta  \right\Vert_{\infty}}_{\text{II}}.
\label{eq:6}
\end{align}
We will upper bound each term on the right-hand side of \eqref{eq:6}. 

\underline{Bounding (I)}: 
Recall that by definition $\bSigma_{\bXhat} \tilde\bbeta = \tilde{\boldeta} =\boldeta\boldeta^\top\widehat{\boldeta}$ i.e., where $\boldeta$ is the eigenvector associated with the largest eigenvalue of $\bLambda = \Var\left\{E(\bXhat \mid y) \right\} = \bGamma^\top \bOmega \bGamma$ with $\bOmega = \Cov\left\{E(\bZ \mid y) \right\}$. Hence, we have $\Vert \hat\boldeta - \bSigma_{\bXhat}\tilde\bbeta\Vert_\infty  =  \Vert \widehat{\boldeta} - \tilde{\boldeta} \Vert_\infty$.  

Following the same notation as in \citet{lin2019sparse}, let $\mcZ_H$ be the $H\times q$ matrix where the $h$th row corresponds to the sample mean of $\bZ$ within the $h$th slice, for $h=1,\ldots, H$, and let $\widehat{\mcX}_H = \mcZ_H \widehat{\bGamma}$ be the corresponding matrix of the sliced mean for $\widehat{\bX}$. Hence, we can project $\widehat{\mcX}_H$ into the column space of $\bLambda$, i.e $\spn(\bGamma^\top \boldeta(\bOmega))$, and decompose
\begin{equation*}
\widehat{\mcX}_H = \mcV_H + \mcT_H,
\end{equation*}
where $\mcV_H = \widehat{\mcX}_H\bP$ and $\mcT_H = \widehat{\mcX}_H (\bI_p - \bP)$, and $\bP = \bGamma^\top \boldeta(\bOmega)\{\boldeta(\bOmega)^{\top} \bGamma \bGamma^\top \boldeta(\bOmega)\}^{-1} \boldeta(\bOmega)^{\top} \bGamma$ is the $p\times p$ projection matrix into the subspace spanned by  $\bGamma^\top \boldeta(\bOmega)$. By the same argument as in \citet{lin2019sparse}, we have
\begin{equation}
\Vert \hat{\boldeta} - \tilde{\boldeta}  \Vert_{\infty} \leq \dfrac{1}{\sqrt{\hat{\lambda}}} \Vert \mcT_H \Vert_{\max},
\label{eq:target}
\end{equation}
where $\hat\lambda$ is the largest eigenvalue for $\widehat{\bLambda} = H^{-1}\widehat{\mcX}_H^\top \widehat{\mcX}_H = H^{-1} \widehat{\bGamma}^\top \mcZ_H^\top \mcZ_H \widehat{\bGamma}$. We next bound $\Vert \mcT_H \Vert_{\max}$. To do it, we also project $\mcZ_H$ into the column space of $\bOmega$, and write
\begin{equation}
\mcZ_H = \mcW_H + \mcU_H,
\end{equation}
where $\mcW_H$ and $\mcU_H$ are similarly, the projection of $\mcZ_H$ to $\spn(\boldeta(\bOmega))$ and its orthogonal complement. Applying the results from \citet{lin2019sparse}, we have $\Vert \mcU_H \Vert_{\max} = O_p\left(\sqrt{\log(q)/n} \right)$. With this result, we obtain
\begin{align}
\mcT_H = \widehat{\mcX}_H(\bI_p-\bP) & = \mcZ_H\widehat{\bGamma}(\bI_p - \bP)  = \mcW_H\widehat{\bGamma}(\bI_p - \bP) + \mcU_H\widehat{\bGamma}(\bI_p - \bP)
\label{eq:13}
\end{align}
We consider each term on the right-hand side of \eqref{eq:13} separately. For the first term, we have
\begin{align}
\mcW_H \widehat{\bGamma}(\bI_p-\bP) = \mcW_H(\widehat{\bGamma} - {\bGamma})(\bI_p - \bP) + \mcW_H{\bGamma}(\bI_p - \bP) \stackrel{(i)}{=} \mcW_H(\widehat{\bGamma} - {\bGamma})(\bI_p - \bP), 
\end{align}
where step $(i)$ follows from the fact that $\mcW_H \in \spn(\boldeta(\bOmega))$ and hence $ \mcW_H{\bGamma}(\bI_p - \bP) =\boldsymbol{0}$. Therefore, we have
\begin{align*}
\Vert \mcW_H \widehat{\bGamma}(\bI_p-\bP) \Vert_{\max} = \Vert \mcW_H(\widehat{\bGamma} - {\bGamma})(\bI_p - \bP)\Vert_{\max} & \leq \Vert \mcW_H \Vert_{\max}\Vert (\widehat{\bGamma} - {\bGamma}) (\bI_p - \bP)\Vert_1  \\ & \leq \Vert \mcW_H \Vert_{\max} \Vert\widehat{\bGamma} - \bGamma\Vert_1 \Vert \bI_p -\bP \Vert_1, \\ & \stackrel{(ii)}{\leq} C\sqrt{\log(q)} \Vert   \mcW_H \Vert_{\max} \Vert\widehat{\bGamma} - \bGamma\Vert_1, 
\end{align*}
where step $(ii)$ follows from the triangle inequality $0 < \Vert \bI_p -\bP \Vert_{1} \leq \Vert \bI_p \Vert_1 + \Vert \bP \Vert _1 = 1+ \Psi_1$. Furthermore, 
$$
\Vert \mcW \Vert_{\max} = \Vert \mcZ_H \boldeta(\bOmega) \boldeta(\bOmega)^{\top} \Vert_{\max} \leq \Vert \mcZ_H \Vert_{\max} \Vert \boldeta(\bOmega) \boldeta(\bOmega)^{\top}  \Vert_1 = O_p(\Psi_2\sqrt{\log q}).
$$
Hence, we have $\Vert \mcW_H \widehat{\bGamma}(\bI_p-\bP) \Vert_{\max} \leq C\Psi_2(1+\Psi_1) \Vert\widehat{\bGamma} - \bGamma\Vert_1$. Using the result from Theorem 3.1, we  obtain
\begin{equation}
\Vert \mcW_H \widehat{\bGamma}(\bI_p-\bP) \Vert_{\max}  \leq C\Psi_2(1+\Psi_1) r\sqrt{\frac{\log pq}{n}}.
\label{eq:15}
\end{equation}

Now, for the second term in \eqref{eq:4}, by a similar argument we have
\begin{equation}
\Vert \mcU_H\widehat{\bGamma}(\bI_p - \bP) \Vert_{\max} \leq \Vert \mcU_H \Vert_{\max} \Vert \widehat{\bGamma} \Vert_{1} \leq C(1+\Psi_1) \sqrt{\dfrac{\log(q)}{n}} \left(\Vert \widehat{\bGamma} - \bGamma \Vert_{1} + \Vert \bGamma\Vert_1   \right),
\end{equation}
where the last step follows from the triangle inequality. Since $\Vert \bGamma \Vert_1 \leq L$, we have
\begin{equation}
\Vert \mcU_H\widehat{\bGamma}(\bI_p - \bP) \Vert_{\max} \leq  C(1+\Psi_1) \left(L \sqrt{\dfrac{\log q}{n}}   +  r\sqrt{\frac{\log q (\log pq)}{n^2}}\right).
\label{eq:17}
\end{equation}

Substituting \eqref{eq:15}  and \eqref{eq:17} into \eqref{eq:13} and applying \eqref{eq:target}, we obtain
\begin{equation}
\Vert \hat{\boldeta} - \tilde{\boldeta}  \Vert_{\infty} \leq \dfrac{1}{\sqrt{\hat{\lambda}}} \Vert \mcT\Vert_{\max} \leq  C(1+\Psi_1) \left(L \sqrt{\dfrac{\log q}{n\hat\lambda}}   + \Psi_2 r\sqrt{\dfrac{\log pq}{n\hat\lambda}} \right), 
\end{equation}
using the assumption that $\log(q) \log(pq) = o(n^2)$. 
By proposition \ref{prop:yonZ}, there exists a positive constant $C$ such that $\lambda(\bOmega)/\hat\lambda(\bOmega) \leq C$. Combining it with the inequality \eqref{eq:weylinequality}, we obtain
$$
\dfrac{\lambda}{\hat\lambda} \leq \dfrac{\lambda(\bOmega)}{\hat\lambda(\bOmega)} \times  \dfrac{\nu_{\max}(\bGamma)}{\nu_{\min}(\widehat{\bGamma})} \leq C,
$$
with a probability of at least $1-\exp\left(-C^\prime \log q \right) - \exp(-C^{\prime\prime} \log pq)$, where the last inequality follows from condition 4.2 and with probability one, the matrix $\widehat{\bGamma}$ obtained from the first stage has full rank.  Hence, with the same probability, we have
$$
\Vert \hat{\boldeta} - \tilde{\boldeta}  \Vert_{\infty} \leq \dfrac{1}{\sqrt{\hat{\lambda}}} \Vert \mcT\Vert_{\max} \leq  C(1+\Psi_1) \left(L \sqrt{\dfrac{\log q}{n\lambda}}   + \Psi_2 r\sqrt{\dfrac{\log pq}{n\lambda}} \right).
$$
\underline{Bounding (II)}: We now bound the second term of \eqref{eq:6}. Using \eqref{eq:maxnormhatsigma} in the proof of Proposition 3, we obtain 
$$
\begin{aligned}
\left\Vert \left(\bSigma_{\bXhat} - \dfrac{1}{n} \hat{\mcX}^\top \hat{\mcX} \right)
\tilde\bbeta  \right\Vert_{\infty} & \leq \Vert \tilde{\bbeta} \Vert_1 \max_{1\leq j,k \leq p} \vert \bgamma_j^\top \bSigma_{\bZ} \bgamma_k - \hat{\bgamma}_j^\top \widehat{\bSigma}_{\bZ}  \hat{\bgamma}_k \vert \\ & \leq CL\left\{L\sqrt{\dfrac{ \log q}{n}} +   \sigma_{\max} r \sqrt{\dfrac{\log pq}{n}} \right\}
\end{aligned}
$$
Finally, with the choice of $\mu_2$ as stated in the theorem, we have 
\begin{equation}
\mu_2 \geq 2  \Vert \hat\boldeta - n^{-1} \hat{\mcX}^\top \hat{\mcX} \tilde\bbeta \Vert_\infty
\label{eq:mu2_theorem 3.2}
\end{equation} with probability at least $1-\exp\left(-C^\prime \log q \right) - \exp(-C^{\prime\prime} \log pq)$. Applying the results from Lemma 1, we finally have
$$
\left\|P(\hat\bbeta) - P({\bbeta})\right\|_F \leq C\frac{\sqrt{s}}{\kappa_2}\left(\rho_1 \sqrt{\frac{\log q}{n}} + \rho_2 r\sqrt{\frac{\log pq}{n}} \right),$$ where $$ \rho_1= L^2 + \dfrac{L(1+ \Psi_1)}{\sqrt{\lambda}}, ~ \rho_2 = L \sigma_{\max} + \dfrac{(1+\Psi_1)\Psi_2}{\sqrt{\lambda}}
$$
with the same probability, as required.

\subsection{Proof of Theorem 3}
The proof of this theorem has the same steps as the primal-dual witness technique, which are used in the proof for variable selection consistency of the Lasso estimator in the regular linear model \citep[see Section 7.5.2,][]{wainwright2019high}, as well as in the two-stage Lasso estimator of \citet{lin2015regularization}. In our context, the proposed two-stage Lasso SIR estimator is the minimizer of 
$$
L(\bbeta) = \dfrac{1}{2n} \Vert \tilde{\by} - \hat{\mcX}\bbeta \Vert _2^2 + \mu_2 \Vert \bbeta \Vert_1.
$$
Applying the Karush--Kuhn-Tucker optimality condition, the estimator $\hat{\bbeta}$ has to satisfy
\begin{equation}
\dfrac{1}{n}\hat{\mcX}^\top_{\hat{S}}(\tilde{\by} - \hat{\mcX}\hat{\bbeta}) = \mu_2 \text{sgn}(\hat{\bbeta}_{\hat{S}}), 
\label{eq: support}
\end{equation}
and 
\begin{equation}
\left\Vert\dfrac{1}{n}\hat{\mcX}^\top_{\hat{S}^c}(\tilde{\by} - \hat{\mcX}\hat{\bbeta}) \right\Vert_\infty \leq \mu_2.
\label{eq: nonsupport}
\end{equation}
It suffices to find a solution $\hat{\bbeta}$ such that both \eqref{eq: support} and \eqref{eq: nonsupport} hold. To achieve this, the idea of the primal-dual witness technique is first let $\hat{\bbeta}_{S^c} = 0$, then construct $\hat{\bbeta}_S$ from \eqref{eq: support}, and finally show that the obtained $\hat{\bbeta}$ satisfies \eqref{eq: nonsupport}.  Central to this development is the following lemma regarding the bound of the sample covariance $\widehat{\bSigma}_{\bXhat}$.

\begin{lemma}
Assume Condition 4.7 is satisfied. If there exists a  constant $C$ such that 
\begin{equation}
CsL\left\{L\sqrt{\dfrac{ \log q}{n}} +   \sigma_{\max} r \sqrt{\dfrac{\log pq}{n}} \right\} \leq \frac{\alpha}{\phi(4-\alpha)}
\label{eq:conditionvariableselection}
\end{equation}
then with probability at least $1-\exp(C^\prime \log q) - \exp(C^{\prime\prime} \log pq)$, the matrix $\widehat{\bSigma}_{\bXhat} = \widehat{\bGamma}^\top \widehat{\bSigma}_{\bZ} \widehat{\bGamma}$, where $\widehat{\bGamma}$ is defined from the first stage with the tuning parameters $\mu_{1,j}$ as stated in Theorem 1, satisfies
\begin{equation}
\left\|\widehat{\bC}_{S S}^{-1}\right\|_{\infty} \leq \frac{4-\alpha}{2(2-\alpha)} \varphi, \quad \left\|\widehat{\bC}_{S^c S}\widehat{\bC}_{S S}^{-1}\right\|_{\infty} \leq 1-\frac{\alpha}{2}.
\label{eq:lemmabound}
\end{equation}
\label{lemmabound}
\end{lemma}

Using a similar argument as in the proof of Theorem 2, 
there also exist constants $C^\prime$ and $C^{\prime\prime}$ such that 
\begin{equation}
\left\Vert\hat{\boldeta}_S  - \widehat{\bC}_{SS} \tilde{\bbeta}_S \right\Vert_\infty \leq \Vert \hat\boldeta - \dfrac{1}{n} \hat{\mcX}^\top \hat{\mcX} \tilde\bbeta \Vert_\infty \leq \dfrac{\alpha}{4-\alpha} \mu_2,
\label{eq:event2}
\end{equation}
with probability at least $1-\exp(-C^\prime \log q) - \exp(-C^{\prime\prime} \log pq)$. Hence, we will condition on both events \eqref{eq:lemmabound} and \eqref{eq:event2} in the remaining part of the proof. 

Next, recall $\widehat{\bC}_{SS} = n^{-1} \hat{\mcX}^\top_S \hat{\mcX}_S$, and by definition of $\tilde{\by}$ we have $n^{-1} \hat{\mcX}^\top \tilde{\by} = \hat{\boldeta}$, which is the eigenvector associated with the largest eigenvalue of $\hat{\bLambda}$. Since $\hat{\bbeta}_{S^c} = 0$, then  \eqref{eq: support} is equivalent to
$$
\begin{aligned}
 \mu_2 \text{sgn}(\hat{\bbeta}_{\hat{S}}) = \dfrac{1}{n}\hat{\mcX}^\top_{\hat{S}}(\tilde{\by} - \hat{\mcX}\hat{\bbeta}) & = \hat{\boldeta}_S  - \dfrac{1}{n}\hat{\mcX}_S^\top \hat{\mcX}_S \hat\bbeta_S  \\
& = \hat{\boldeta}_S  - \widehat{\bC}_{SS} (\hat\bbeta_S - \tilde{\bbeta}_S) - \widehat{\bC}_{SS} \tilde{\bbeta}_S \\
& = (\hat{\boldeta}_S  -\widehat{\bC}_{SS} \tilde{\bbeta}_S) -\widehat{\bC}_{SS} (\hat\bbeta_S - \tilde{\bbeta}_S),
\end{aligned}
$$
where $\tilde{\bbeta}$ is defined previously. Hence, we obtain
\begin{equation}
\hat\bbeta_S - \tilde{\bbeta}_S = \widehat{\bC}_{SS}^{-1} \left\{(\hat{\boldeta}_S  - \widehat{\bC}_{SS} \tilde{\bbeta}_S) - \mu_2 \text{sgn}(\hat{\bbeta}_{\hat{S}})   \right\}, 
\label{eq:formula_diff_betaS}
\end{equation}
and therefore
\begin{equation}
\Vert \hat\bbeta_S - \tilde{\bbeta}_S  \Vert_\infty \leq \Vert \widehat{\bC}_{SS}^{-1} \Vert_\infty \left\{ \left\Vert\hat{\boldeta}_S  - \widehat{\bC}_{SS} \tilde{\bbeta}_S \right\Vert_\infty + \mu_2  \right\}.
\label{eq:diff_betaS}
\end{equation}
 Substituting \eqref{eq:lemmabound} and \eqref{eq:event2} into \eqref{eq:diff_betaS}, we obtain
\[
\Vert \hat\bbeta_S - \tilde{\bbeta}_S  \Vert_\infty \leq \frac{4-\alpha}{2(2-\alpha)} \varphi \left( \dfrac{\alpha}{4-\alpha}\mu_2  + \mu_2 \right) = \dfrac{2}{(2-\alpha)} \varphi\mu_2 < \min_{j \in S}\vert \tilde\beta_j\vert,
\]

where the last step follows from 
$$
\min_{j \in S}\vert \tilde\beta_j\vert = k^{-1} \vert \boldeta^\top \widehat{\boldeta} \vert \min_{j \in S}\vert \beta_j\vert > k^{-1} \hat{\rho} \min_{j \in S}\vert \beta_j\vert > \dfrac{2}{(2-\alpha)} \varphi\mu_2,
$$
and by the choice of $\mu_2$ as stated in the theorem. Thus  we have $\text{sgn}(\hat\bbeta_S) = \text{sgn}(\tilde{\bbeta}_{S})$, and since $\hat\bbeta_{S^c} = 0$ by construction then we have $\hat{S} = S$. 

It remains to show that $\hat{\bbeta}$ satisfies \eqref{eq: nonsupport}. We have
\[
\begin{aligned}
\dfrac{1}{n}\hat{\mcX}^\top_{\hat{S}^c}(\tilde{\by} - \hat{\mcX}\hat{\bbeta}) & = \hat{\boldeta}_{S^c} - \dfrac{1}{n}\hat{\mcX}_{S^c}^{\top}\hat{\mcX}_S\hat{\bbeta}_S \\
& = \hat{\boldeta}_{S^c}- \dfrac{1}{n}\hat{\mcX}^{\top}_{S^c}\hat{\mcX}_S(\hat{\bbeta}_S - \tilde{\bbeta}_S) - \dfrac{1}{n}\hat{\mcX}_{S^c}^{\top}\hat{\mcX}_S \tilde{\bbeta}_S  \\
& = \left(\hat{\boldeta}_{S^c} - \dfrac{1}{n}\hat{\mcX}_{S^c}^{\top}\hat{\mcX}_S \tilde{\bbeta}_S \right) - \widehat{\bC}_{S^c S} \widehat{\bC}_{SS}^{-1} \left\{(\hat{\boldeta}_S  - \widehat{\bC}_{SS} \tilde{\bbeta}_S) - \mu_2 \text{sgn}(\hat{\bbeta}_{{S}}) \right\},  
\end{aligned}
\]
where the last equality follows from \eqref{eq:formula_diff_betaS}. Hence, we obtain
\begin{align*}
\left\Vert\dfrac{1}{n}\hat{\mcX}^\top_{\hat{S}^c}(\tilde{\by} - \hat{\mcX}\hat{\bbeta})\right\Vert_\infty & \leq \left\Vert \hat{\boldeta}_{S^c} - \dfrac{1}{n}\hat{\mcX}_{S^c}^{\top}\hat{\mcX}_S \tilde{\bbeta}_S\right\Vert_\infty + \Vert \widehat{\bC}_{S^c S} \widehat{\bC}_{SS}^{-1} \Vert_\infty \Vert (\hat{\boldeta}_S  - \widehat{\bC}_{SS} \tilde{\bbeta}_S) - \mu_2 \text{sgn}(\hat{\bbeta}_{{S}}) \Vert_\infty
\\ & \leq \left\Vert \hat{\boldeta}- \dfrac{1}{n}\hat{\mcX}^{\top}\hat{\mcX} \tilde{\bbeta}\right\Vert_\infty + \Vert \widehat{\bC}_{S^c S} \widehat{\bC}_{SS}^{-1} \Vert_\infty \left\Vert (\hat{\boldeta}_S - \dfrac{1}{n} \widehat{\mcX}^{\top}\widehat{\mcX}  \tilde{\bbeta}) - \mu_2 \text{sgn}(\hat{\bbeta}_{{S}}) \right\Vert_\infty \\
& \leq \dfrac{\alpha}{4-\alpha} \mu_2 + \left(1-\frac{\alpha}{2}\right)\left(\dfrac{\alpha}{4-\alpha} \mu_2 + \mu_2 \right) = \mu_2.
\end{align*}
The required results follow from this inequality.

\subsection{Proof of Theorem 4}
Consider the multiple index model with $\bB \in \mathbb{R}^{p \times d}$. In this case, the quantities $\boldeta$ and $\widehat{\boldeta}$ are the $p \times d$ semi-orthogonal matrices of eigenvectors of $\bLambda$ and $\widehat{\bLambda}$, respectively. Let $\widetilde{\boldeta} = P(\boldeta)\widehat{\boldeta} = \boldeta\boldeta^\top\widehat{\boldeta}$ be the projection of $\widehat{\boldeta}$ on $\boldeta$. By Lemma 1, we have $\bSigma_{\bXhat}\bB= \boldeta\widetilde{\bD}$ where $\widetilde{\bD}$ is a non-zero $d \times d$ diagonal matrix, such that $\boldeta = \Sigma_{\bXhat}\bB\widetilde{\bD}^{-1}$. Hence, we can write
$$
\widetilde{\boldeta} = \Sigma_{\bXhat} \bB\widetilde{\bD}^{-1}\boldeta^\top{\tilde\boldeta} = \bSigma_{\bXhat} \widetilde{\bB},
$$
where $\widetilde{\bB} =\widetilde{\bD}^{-1}\boldeta^\top{\widehat{\boldeta}}$. When $\bH= \boldeta^\top{\widehat{\boldeta}}$ is a $d \times d$ invertible matrix, the column space spanned by $\tilde{\bB}$ and that spanned by $\bB$ are the same, and so are the index sets for their non-zero rows. 

Proposition \ref{prop:propS3} establishes a lower bound on the trace of $\bH^2$.

\begin{proposition}
Assume Conditions 4.1--4.6 are satisfied. Then there exist positive constants $C$,  $C^\prime$, and $C^{\prime\prime}$ such that
$$
\tr(\bH^2) \geq d - \dfrac{CL\sqrt{d}}{\lambda_d}\left\{L\sqrt{\dfrac{ \log q}{n}} +   \sigma_{\max} r \sqrt{\dfrac{\log pq}{n}} \right\}, 
$$
with probability at least $1-\exp(C^\prime \log q) - \exp(C^{\prime\prime} \log pq)$. 
\label{prop:propS3}
\end{proposition}
Furthermore, by the Cauchy-Schwartz inequality, we have $\tr(\bH^2) \leq \left\{\tr(\bH) \right\}^2 \leq d$. By Condition 4.3 with $r^2 \log(pq)/n \to 0$ and  Condition 4.4 with $\lambda_d = O(1)$, we thus have $\tr(\bH^2) \to d$ with a probability at least $1-\exp(C^\prime \log q) - \exp(C^{\prime\prime} \log pq)$. This implies that with the same probability, the matrix $\bH$ is invertible. 

Conditioning on the invertibility of $\bH$, Lemma 3 holds for each column $\tilde{\bbeta}_k, ~k = 1,\ldots, d$. As a result, we have
$$
\norm{\hat\bbeta_k -\tilde{\bbeta}_k}_2 \leq \dfrac{3}{\kappa_2} \mu_{2k} \sqrt{s},
$$
for any $\mu_{2k} \geq 2 \norm{\boldeta_k - n^{-1} \widehat{\mcX}^\top \widehat{\mcX} \widetilde{\bbeta}_k}_{\infty}$. Hence, by a similar argument to the proof of Theorem 2, and with the choice of $\mu_2$ as stated in the theorem, there exist constants $C, C^\prime$, and $C^{\prime\prime}$ such that
$$
\left\|P(\hat\bbeta_k) - P({\bbeta_k})\right\|_F \leq C\frac{\sqrt{s}}{\kappa_2}\left(\rho_1 \sqrt{\frac{\log q}{n}} + \rho_2 r\sqrt{\frac{\log pq}{n}} \right),$$ where $$ \rho_1= L^2 + \dfrac{L(1+ \Psi_1)}{\sqrt{\lambda_d}}, ~ \rho_2 = L \sigma_{\max} + \dfrac{(1+\Psi_1)\Psi_2}{\sqrt{\lambda}_d}
$$
with probability at least $1-\exp(C^\prime \log q) - \exp(C^{\prime\prime} \log pq)$. Hence, by the same argument as in the proof of Theorem 2 in \citet{lin2019sparse}, we also have 
$$
\Vert P(\widehat{\bB}) - P(\bB) \Vert_F \leq C\frac{\sqrt{s}}{\kappa_2}\left(\rho_1 \sqrt{\frac{\log q}{n}} + \rho_2 r\sqrt{\frac{\log pq}{n}} \right),
$$
with probability at least  $1-d\exp(C^\prime \log q) - d\exp(C^{\prime\prime} \log pq)$   as claimed. 

Finally, turning to variable selection consistency, let $\mathcal{S}_k = \left\{k: \beta_{jk} \neq 0 \right\}$ and $\widehat{\mathcal{S}}_k = \left\{k: \hat{\beta}_{jk} \neq 0 \right\}$.  By a similar argument to that in the proof of Theorem 3, under the condition on $\mu_k$ as stated in Theorem 4, we have $\widehat{\mathcal{S}}_k = \mathcal{S}_k $ for $k=1,\ldots, d$ with probability at least $1-\exp(C^\prime \log q) - \exp(C^{\prime\prime} \log pq)$. Note the diagonal element of the estimated projection matrix $[\mathcal{P}(\widehat{\bB})]_{jj} = 0$ if and only if all one $\hat{\beta}_{jk} = 0$ across $k=1,\ldots, d$. Hence, taking the union bound, we have $\mathcal{S} = \mathcal{S}$ with a probability at least $1-d\exp(C^\prime \log q) - d\exp(C^{\prime\prime} \log pq)$.

\section{Additional derivations}
\label{sec:additionalproofs}
\subsection{Proof of Lemma \ref{lemma:boundingnormmax}}
\label{section:boundingsecondterm}

By the triangle inequality, we have
$$
\norm{\widehat{\bOmega} - \bOmega}_{\max} = \norm{(\bSigma_\bZ - \widehat{\bSigma}_{\bZ}) + (\bXi - \widehat{\bXi})}_{\max} \leq \norm{\widehat{\bSigma}_{\bZ} - \bSigma_{\bZ}}_{\max} + \norm{\widehat{\bXi} - \bXi}_{\max}.
$$
Also, by the property of sub-Gaussian random variables, there exist positive constants $C$ and $C^\prime$ such that
\begin{equation}
\left\|\widehat{\bSigma}_{\bZ}-\bSigma_{\bZ}\right\|_{\max }=C\sqrt{\frac{\log q}{n}},
\label{eq:estSigmaZ}
\end{equation}
with probability at least $1-\exp(-C^\prime \log q)$ \citep{ravikumar2011high, tan2018convex}. Next, we will bound the elementwise maximum norm $\norm{\widehat{\bXi} - \bXi}_{\max}.$

Following \citet{zhu2006sliced}, we introduce the following notations. Let $y_{(1)}, \ldots, y_{(n)}$ denote the order statistics of the outcome data $y_1, \ldots, y_n$, and ${\bZ}_{(1)}, \ldots, \bZ_{(n)}$ be the concomitants in the $\mcZ$ data. Define $m(y_{(i)}) = E(\bZ_{(i)} \mid y_{(i)}) \in \mathbb{R}^{q}$ and $\bm\epsilon_{(i)} = \bZ_{(i)} - m(y_{(i)})$, and let $\bM_n$ be the $n \times p$ matrix whose $i$th row is $m(y_{(i)})$ and $\bE_n$  be the $n \times p$ matrix whose $i$th row is $\bm{\epsilon}_i$. Note we have $E(\bm\epsilon_{(i)}) = 0$ and $\Cov\left\{m(y_{(i)}), \bm\epsilon_{(i)}\right\} = 0$ for all $i=1,\ldots,n$. 

Let $Z_{(i)j}$, $m_j(y_{(i)})$ and $\epsilon_{(i)j}$ be the $j$th element of $\bZ_{(i)}$, $m(y_{(i)})$ and $\epsilon_{(i)}$, respectively, for $j=1,\ldots, q$. Similar to Proposition 2 in the Supplementary Materials of \citet{tan2018convex}, we obtain the following properties of $m(y_{i})$ and $\epsilon_{(i)}$, whose proof is omitted.  

\begin{proposition}
Under Condition 4.1, $Z_{(i)j}$, $m_j(y_{(i)})$ and $\epsilon_{(i)j}$ are sub-Gaussian with sub-Gaussian norms $\norm{\bZ_j}_{\psi_2} \leq C\sqrt{(\bSigma_{\bZ})_{jj}}$, $\norm{\bZ_j}_{\psi_2}$ and $2\norm{\bZ_j}_{\psi_2}$, respectively, for $i=1,\ldots,n, ~j=1,\ldots, q$.
\label{prop:subGaussianerror}
\end{proposition}
Noting $n = cH$ and using the same decomposition as in \citet{zhu2006sliced}, write
\begin{equation}
\begin{aligned}
\widehat{\bXi} & = \frac{1}{H(c-1)}\left\{\left(1-\frac{1}{c}\right) \bE_n^\top \bE_n\right\}+\frac{1}{H(c-1)}\left\{\bE_n^\top\left(\bI_H \otimes \bQ_c\right) \bE_n\right\}
\\ & +\frac{1}{H(c-1)}\left\{\bE_n^\top\left(\bI_H \otimes \bP_c\right) \bM_n+\bM_n^\top\left(\bI_H \otimes \bP_c\right) \bE_n\right\} 
+\frac{1}{H(c-1)} \bM_n^\top\left(\bI_H \otimes \bP_c\right) \bM_n \\ 
& := \bJ_1 + \bJ_2 + \bJ_3 + \bJ_4,
\end{aligned}
\end{equation}
where $\otimes$ denotes the Kronecker product, $\bP_c = \bI_c - c^{-1} \mathbf{e}_c\mathbf{e}_c^\top$, $\bQ_c = c^{-1}(\bI_c - \mathbf{e}_c\mathbf{e}_c^\top)$, and $\mathbf{e}_c$ is the $c\times 1$ vector of ones. By the triangle inequality, we obtain
\begin{equation}
 \Vert \bXi - \widehat{\bXi} \Vert_{\max}  \leq \Vert \bJ_1 - {\bXi} \Vert_{\max} + \norm{\bJ_2}_{\max} + \norm{\bJ_3}_{\max} + \norm{\bJ_4}_{\max}.
\label{eq:XiminusXihat}
\end{equation}
We proceed by bounding each term on the right hand side of \eqref{eq:XiminusXihat}. 

For the first term $\Vert \bJ_1 - {\bXi} \Vert_{\max}$, noting that $\bJ_1 = n^{-1}\bE_n^\top \bE_n = n^{-1}\sum_{i=1}^{n} \bm\epsilon_i\bm{\epsilon}_i^\top $, then $\text{E}(\bJ_1) = \bXi$. On combining this with Proposition \ref{prop:subGaussianerror}, it follows that each element of $\bJ_1 - \bXi$ is a centered sub-exponential random variable with sub-exponential norm bounded by $CL^2$, since $\norm{\bZ}_{\psi_2} \leq L$ by condition 1. Therefore, with probability at least $1-\exp(-C^\prime \log q)$, we have
\begin{equation}
\Vert \bJ_1 - {\bXi} \Vert_{\max} \leq C \sqrt{\frac{\log q}{n}}. 
\label{eq:termJ1}
\end{equation}
For term $\norm{\bJ_2}_{\max}$, we note that $\bQ_c$ has all diagonal elements zero and off-diagonal elements $-c^{-1}$, so each element of $\bJ_2$ is the sum of $n$ terms of the form $[H(c-1)]^{-1}\epsilon_{ij}\epsilon_{i^\prime j^\prime}$ with $i\neq i^\prime$ (note $j$ and $j^\prime$ can be the same). Next, conditional on the order statistics $y_{(1)}, \ldots, y_{(n)}$, the quantities $\bm{\epsilon}_{i}$ and $\bm{\epsilon}_{i^\prime}$ are independent \citep{yang1977general}. On combining this with Proposition \ref{prop:subGaussianerror}, we have that $\epsilon_{ij}\epsilon_{i^\prime j^\prime}$ is zero-mean sub-exponential with sub-exponential norm bounded by $CL^2$. Therefore, with probability at least $1-\exp(-C^\prime \log q)$, we have
\begin{equation}
\Vert \bJ_2 \Vert_{\max} \leq C \sqrt{\frac{\log q}{n}}. 
\label{eq:termJ2}
\end{equation}
Next, we consider the term $\bJ_3$, where it suffices to consider the term $\bE_n^\top (\bI_H \otimes \bP_c) \bM_n$. Let $\widetilde{\bM}_n$ denote the $n \times p$ matrix whose $i$th row is given by $m(y_{(i)}) - m(y_{(i+1)})$ for $1\leq i \leq n-1$, and whose $n$th row is $\bm{y}_{(n)}$. It is straightforward to see that $\widetilde{\bM}_n = \bA_n \bM_n$ and $\bM_n = \bA_n^{-1}\widetilde{\bM}_n  $, where for any $t$, we have
\[
\bA_t=\left(\begin{array}{cccc}
1 & -1 & & \\
& 1 & \ddots & \\
& & \ddots & -1 \\
& & & 1
\end{array}\right)_{t\times t} ~\bA_t^{-1}=\left(\begin{array}{cccc}
1 & 1 & \ldots & 1 \\
& 1 & \ldots & \vdots \\
& & \ddots & 1 \\
& & & 1
\end{array}\right)_{t \times t}.
\]
Hence, by the property of the Kronecker product, we obtain
\begin{align*}
\norm{\bE_n^\top\left(\bI_H \otimes \bP_c\right) \bM_n}_{\max}& =\norm{\bE_n^\top \left(\bI_H \otimes \bP_c\right) \bA_n^{-1} \widetilde{\bM}_n^T}_{\max} \\ & =\norm{\bE_n^\top \left\{\bI_H \otimes\left(\bP_c \bA_c^{-1}\right)\right\} \widetilde{\bM}_n}_{\max} \\
& \leq \norm{\bE_n}_{\max} \norm{\left\{\bI_H \otimes\left(\bP_c \bA_c^{-1}\right)\right\} \widetilde{\bM}_n}_1.
\end{align*}
We have $\norm{\bE_n}_{\max} = \max_{i,j} \vert \epsilon_{ij}\vert$, where all $\epsilon_{i,j}$ are sub-Gaussian random variables by Proposition \ref{prop:subGaussianerror} for all $i=1,\ldots,n$ and $j=1,\ldots, q$. Therefore, with probability at least  $1-\exp(C^\prime\log q)$, we have that $\norm{\bE_n}_{\max} \leq C \sqrt{\log d}$. Also note that in the matrix $\bI_H \otimes\left(\bP_c \bA_c\right)$, all the columns of  $\bI_H \otimes\left(\bP_c \bA_c\right)$ 
have their sum of absolute values less than two, i.e $\norm{\bI_H \otimes\left(\bP_c \bA_c\right)}_1 \leq 2$. On combining these results, we obtain
\begin{equation*}
\norm{\left\{\bI_H \otimes\left(\bP_c \bA_c\right)\right\} \widetilde{\bM}_n}_{1} \leq 2\norm{\widetilde{\bM}_n}_{1} \leq \sum_{i=2}^n\norm{m\left(y_{(i)}\right)-m\left(y_{(i-1)}\right)}_{\infty} \leq C n^{1/4},
\end{equation*}
where the last inequality follows from condition 4.6.  Therefore, with probability at least $1-\exp(C^\prime\log q)$, we obtain
\begin{equation}
\norm{\bJ_3}_{\max} \leq C \left(\dfrac{1}{n} \sqrt{\log q} ~  n^{1/4} \right) = \dfrac{C}{n^{1/4}} \sqrt{\frac{\log q}{n}}.
\label{eq:termJ3}
\end{equation}

Finally, we consider the term $\bJ_4$. Substituting $\bM_n = \bA_n^{-1}\widetilde{\bM}_n$ and using properties of Kronecker products, we have
$$
\norm{\bM_n^\top \left(\bI_H \otimes \bP_c\right)\bM_n}_{\max} = \norm{\widetilde{\bM}_n^\top\left(\bI_H \otimes \bA_c^{-\top} \bP_c \bA_c^{-1}\right) \widetilde{\bM}_n}_{\max} .
$$
Since all the elements of $\bI_H \otimes \bA_c^{-\top} \bP_c \bA_c^{-1}$ have absolute value less than $(c/4)$, then 
\begin{equation}
\norm{\bJ_4}_{\max} \leq \dfrac{C}{n} \norm{\widetilde{\bM}_n}_{1}^2  \leq \dfrac{C}{n} \left[\sup_{\Pi_n(B)} \sum_{i=2}^n\left\|m\left\{y_{(i)}\right\}-m\left\{y_{(i-1)}\right\}\right\|_{\infty}\right]^2 \leq \dfrac{C}{\sqrt{n}},
\label{eq:termJ4}
\end{equation}
where the last inequality follows from Condition 4.6. 

Substituting \eqref{eq:termJ1}-\eqref{eq:termJ4} into \eqref{eq:XiminusXihat}, we obtain  
\begin{equation}
\norm{\bXi - \widehat{\bXi}}_{\max} \leq C \sqrt{\frac{\log q}{n}},
\end{equation}
with probability at least $1-\exp(-C^\prime \log q)$. Hence, with the same probability we obtain
$$\norm{\widehat{\bOmega}- \bOmega}_{\max} \leq C \sqrt{\frac{\log q}{n}},
$$
as required.

\subsection{Proof of Proposition \ref{prop:Re2}}
\label{proof_prop:Re2}
Let $\bdelta \in \mathcal{C}(S, 3)$. Then we have
\begin{equation}
\begin{aligned}
\norm{\widehat{\bSigma}^{1/2}_{\bXhat} \bdelta}_2^2 = \bdelta^\top \widehat{\bSigma}_{\bXhat} \bdelta = \bdelta^\top (\widehat{\bSigma}_{\bXhat} - \bSigma_{\bXhat})\bdelta + \bdelta^\top \bSigma_{\bXhat}\delta 
 & \geq \kappa_1 \norm{\bdelta}_2^2 -  \bdelta^\top (\widehat{\bSigma}_{\bXhat} - \bSigma_{\bXhat})\bdelta, 
\end{aligned}
\label{eq:REforhatSigmaX}
\end{equation}
where the last inequality follows from Condition 4.5 on the restricted eigenvalue of $\bSigma_{\bXhat}$. Next, we provide an upper bound on the second term. We have
$$
\bdelta^\top (\widehat{\bSigma}_{\bXhat} - \bSigma_{\bXhat})\bdelta  \leq \norm{\bdelta}_1^2 \norm{\widehat{\bSigma}_{\bXhat} - \bSigma_{\bXhat}}_{\max}.
$$

Since $\bdelta \in \mathcal{C}(S, 3)$, i.e $\norm{\bdelta_{S^c}}_1 \leq 3\norm{\bdelta_{S}}_1$, then we obtain $\norm{\bdelta}_1^2 = (\norm{\bdelta_{S}}_1 + \norm{\bdelta_{S^c}}_1)^2 \leq 16 \norm{\bdelta_{S}}_1^2 \leq 16s \norm{\bdelta}_2^2$. Furthermore, noting that $\bSigma_{\bXhat} = \bGamma^\top \bSigma_\bZ \bGamma$ and $\widehat{\bSigma}_{\bXhat} = \widehat{\bGamma}^\top \widehat{\bSigma}_\bZ \widehat{\bGamma}$, then we have
$$
\norm{\widehat{\bSigma}_{\bXhat} - \bSigma_{\bXhat}}_{\max} = \max_{j,k} \vert \bgamma_j^\top \bSigma_{\bZ} \bgamma_k - \hat{\bgamma}_j^\top \widehat{\bSigma}_{\bZ}  \hat{\bgamma}_k \vert.
$$

By the triangle inequality, we obtain
\begin{equation}
\begin{aligned}
\vert \bgamma_j^\top \bSigma_{\bZ} \bgamma_k - \hat{\bgamma}_j^\top \widehat{\bSigma}_{\bZ}  \hat{\bgamma}_k \vert  & =   \vert (\bgamma_j^\top \bSigma_{\bZ} \bgamma_k  - \hat{\bgamma}_j^\top \bSigma_{\bZ} \bgamma_k)  + (\hat{\bgamma}_j^\top \bSigma_{\bZ} \bgamma_k - \hat{\bgamma}_j^\top \widehat{\bSigma}_{\bZ} \bgamma_k)  + (\hat{\bgamma}_j^\top \widehat{\bSigma}_{\bZ} \bgamma_k  - \hat{\bgamma}_j^\top \widehat{\bSigma}_{\bZ} \hat{\bgamma}_k) \vert \\
& \leq \vert \bgamma_j^\top \bSigma_{\bZ} \bgamma_k  - \hat{\bgamma}_j^\top \bSigma_{\bZ} \bgamma_k \vert + \vert \hat{\bgamma}_j^\top \bSigma_{\bZ} \bgamma_k - \hat{\bgamma}_j^\top \widehat{\bSigma}_{\bZ} \bgamma_k \vert + \vert \hat{\bgamma}_j^\top \widehat{\bSigma}_{\bZ} \bgamma_k  - \hat{\bgamma}_j^\top \widehat{\bSigma}_{\bZ} \hat{\bgamma}_k \vert \\
& \leq \Vert \hat{\bgamma}_j-\bgamma_j \Vert_1 \Vert \bSigma_{\bZ}\bgamma_k \Vert_\infty + \Vert \hat\bgamma_j \Vert_1 \Vert \bSigma_\bZ - \widehat{\bSigma}_\bZ \Vert_{\max} \Vert \bgamma_k \Vert_1 +  \Vert \widehat{\bSigma}_\bZ \hat\bgamma_j  \Vert_\infty \Vert  \hat{\bgamma}_k-\bgamma_k \Vert_1  \\& = \tilde{T}_1 + \tilde{T}_2 + \tilde{T}_3.
\end{aligned}
\label{eq:sumterm}
\end{equation}

The three terms $\tilde{T}_1$,  $\tilde{T}_2$ and $\tilde{T}_3$ are similar to $T_1, T_2$ and $T_3$ in Section \ref{section:hatbLambda} with $\bOmega$ and $\widehat{\bOmega}$ replaced by $\bSigma_\bZ$ and $\widehat{\bSigma}_{\bZ}$, respectively. It follows that by a similar argument, we obtain
\begin{equation}
\norm{\widehat{\bSigma}_{\bXhat} - \bSigma_{\bXhat}}_{\max} = \max_{j,k}   \vert \bgamma_j^\top \widehat{\bSigma}_{\bXhat} \bgamma_k - \hat{\bgamma}_j^\top \bSigma_{\bXhat}\hat{\bgamma}_k \vert \leq CL\left\{L\sqrt{\dfrac{ \log q}{n}} +   \sigma_{\max} r \sqrt{\dfrac{\log pq}{n}} \right\}, 
\label{eq:maxnormhatsigma}
\end{equation}
with probability at least $1-\exp(-C^\prime \log q) -\exp(-C^{\prime\prime} \log pq)$.

Substituting the above into \eqref{eq:REforhatSigmaX} and under the assumption $L_1^2\sigma_{\max}rs\sqrt{(\log pq)/{n}} = o(1)$, then there exists a positive constant $0 < \kappa_2 < \kappa_1$ such that 
$$
\norm{\widehat{\bSigma}^{1/2}_{\bXhat} \bdelta}_2^2 \geq \kappa_2 \norm{\bdelta}_2^2,
$$
with probability at least $1- \exp(-C^\prime \log q) - \exp(-C^{\prime\prime} \log pq)$, as required. 

\subsection{Proof of Lemma \ref{lemmabound}}
From equation \eqref{eq:maxnormhatsigma} in the proof of Proposition 3, with probability at least  $1- \exp(-Cq) - \exp(-C^\prime \log pq)$ we have
\begin{align*}
\norm{\widehat{\bSigma}_{\bXhat} - \bSigma_{\bXhat}}_{\max} = \max_{j,k}   \vert \bgamma_j^\top \widehat{\bSigma}_{\bXhat} \bgamma_k - \hat{\bgamma}_j^\top \bSigma_{\bXhat}\hat{\bgamma}_k \vert \leq CL\left\{L\sqrt{\dfrac{ \log q}{n}} +   \sigma_{\max} r \sqrt{\dfrac{\log pq}{n}} \right\}. \end{align*}
Conditioning on this event, and combining it with  \eqref{eq:conditionvariableselection}, we obtain
$$
\varphi \| \widehat{\bC}_{SS} - \bC_{SS} \|_\infty \leq C\phi s L\left\{L\sqrt{\dfrac{ \log q}{n}} +   \sigma_{\max} r \sqrt{\dfrac{\log pq}{n}} \right\} \leq \frac{\alpha}{4 - \alpha},
$$ 
and similarly, 
$$\varphi \| \widehat{\bC}_{S^cS} - \bC_{S^cS} \|_\infty \leq \frac{\alpha}{4 - \alpha}.
$$

Next, by an error bound for matrix inversion \citep[~Section 5.8]{horn2012matrix}, we obtain
$$
\| (\widehat{\bC}_{SS})^{-1} - (\bC_{SS})^{-1} \|_\infty \leq \frac{\varphi \| \widehat{\bC}_{SS} - \bC_{SS} \|_\infty}{1 - \varphi \| \widehat{\bC}_{SS} - \bC_{SS} \|_\infty} \varphi \leq \frac{\alpha}{2(2 - \alpha) \varphi}.
$$
Applying the triangle inequality, we have
$$
\| (\widehat{\bC}_{SS})^{-1} \|_\infty \leq \| (\bC_{SS})^{-1} \|_\infty + \| (\widehat{\bC}_{SS})^{-1} - (\bC_{SS})^{-1} \|_\infty \leq \varphi + \frac{\alpha}{2(2 - \alpha) \varphi} = \frac{4 - \alpha}{2(2 - \alpha) \varphi},
$$
which proves the first inequality in \eqref{eq:lemmabound}. 

To show the second inequality, we can write
$$
\widehat{\bC}_{SS} (\widehat{\bC}_{SS})^{-1} - \bC_{SS} (\bC_{SS})^{-1} = (\widehat{\bC}_{SS} - \bC_{SS}) (\widehat{\bC}_{SS})^{-1} - \bC_{SS} (\widehat{\bC}_{SS})^{-1} (\widehat{\bC}_{SS} - \bC_{SS}) (\bC_{SS})^{-1}.
$$
It follows that
\begin{align*}
& \| \widehat{\bC}_{S^cS} (\widehat{\bC}_{SS})^{-1} - \bC_{S^cS} (\bC_{SS})^{-1} \|_\infty \\[1em] &  \leq \| \widehat{\bC}_{S^cS} - \bC_{S^cS} \|_\infty \| (\widehat{\bC}_{SS})^{-1} \|_\infty + \| \bC_{S^cS} (\bC_{SS})^{-1} \|_\infty \| \widehat{\bC}_{SS} - \bC_{SS} \|_\infty \| (\widehat{\bC}_{SS})^{-1} \|_\infty
\\[1em] & \leq \frac{\alpha} {(4 - \alpha) \varphi} \times  \dfrac{4-\alpha}{2(2 - \alpha)} \varphi + (1 - \alpha) \times \frac{\alpha}{(4 - \alpha) \varphi} \times \dfrac{4-\alpha}{2(2 - \alpha)} \varphi  = \frac{\alpha}{2},
\end{align*}
as required.

\subsection{Proof of Proposition \ref{prop:propS3}}

The proof of this proposition generalizes the proof of Lemma 2 above. By the variational definition of eigenvectors, $\hat{\boldeta}$ maximizes $\tr(\boldv^\top \hat{\bLambda} \boldv)$ subject to $\boldv^\top\boldv=\bI_d$, and so
$$
\tr(\boldeta^\top \hat{\bLambda}\boldeta) \leq \tr(\hat{\boldeta}^\top \hat{\bLambda}\hat{\boldeta}), 
$$
or equivalently $\langle \hat\bLambda, \boldeta\boldeta^\top \rangle \leq \langle \hat{\bLambda}, \hat{\boldeta}\hat{\boldeta}^\top \rangle$, where for two matrices $\bA_1$ and  $\bA_2$ we let $\langle \bA_1, \bA_2 \rangle = \tr(\bA_1^\top \bA_2)$ denote their dot product.  Writing $\hat\bLambda = \bLambda + (\hat\bLambda - \bLambda)$, it follows that
\begin{equation}
\langle\bLambda, \boldeta\boldeta^\top - \hat{\boldeta}\hat{\boldeta}^\top \rangle \leq \langle \hat\bLambda - \bLambda,  \boldeta\boldeta^\top - \hat{\boldeta}\hat{\boldeta}^\top \rangle.
\label{eq:basicInequality2}
\end{equation}
For the left-hand side of \eqref{eq:basicInequality2}, by an eigendecomposition we have
$
\bLambda = \boldeta \text{diag}(\lambda_1, \ldots, \lambda_d) \boldeta^\top$ and $\boldeta^\top \boldeta = \bI_d$. It follows that
$$
\begin{aligned}
\langle\bLambda, \boldeta\boldeta^\top - \hat{\boldeta}\hat{\boldeta}^\top \rangle &  = \tr(\bLambda) - \tr\{\boldeta \text{diag}(\lambda_1, \ldots, \lambda_d) \boldeta^\top \widehat{\boldeta}
\widehat{\boldeta}^\top\}\\ & = \tr(\bLambda) - \tr\{\text{diag}(\lambda_1, \ldots, \lambda_d)\bH^2\} \\
& = \sum_{k=1}^{d} \lambda_k(1-h^2_{kk}) \geq \lambda_d (d-\tr(\bH^2). 
\end{aligned}
$$ 

For the right hand side of \eqref{eq:basicInequality2}, applying H\"{o}lder's inequality we have
\[
\begin{aligned}
\langle \hat\bLambda - \bLambda,  \boldeta\boldeta^\top - \hat{\boldeta}\hat{\boldeta}^\top \rangle & \leq \norm{\hat\bLambda - \bLambda}_{\max} \norm{\boldeta\boldeta^\top - \hat{\boldeta}\hat{\boldeta}^\top}_1 \\ & \leq \sqrt{d} \norm{\hat\bLambda - \bLambda}_{\max}  \\ & \leq CL\sqrt{d}\left(L\sqrt{\dfrac{ \log q}{n}} +   \sigma_{\max} r \sqrt{\dfrac{\log pq}{n}} \right), 
\end{aligned}
\]
where the last inequality follows from  Lemma \ref{lemma:boundingnormmax} above . Hence, we have
$$
\lambda_d \left\{d-\tr(\bH^2)\right\} \leq CL\sqrt{d}\left(L\sqrt{\dfrac{ \log q}{n}} +   \sigma_{\max} r \sqrt{\dfrac{\log pq}{n}} \right), 
$$
or equivalently
$$
\tr(\bH^2) \geq d - \dfrac{CL\sqrt{d}}{\lambda_d}\left(L\sqrt{\dfrac{ \log q}{n}} +   \sigma_{\max} r \sqrt{\dfrac{\log pq}{n}} \right),
$$
as required.

\FloatBarrier
\section{Additional results for the application}
\subsection{Mouse obesity data}
\begin{figure}[ht]
\includegraphics[width = \textwidth]{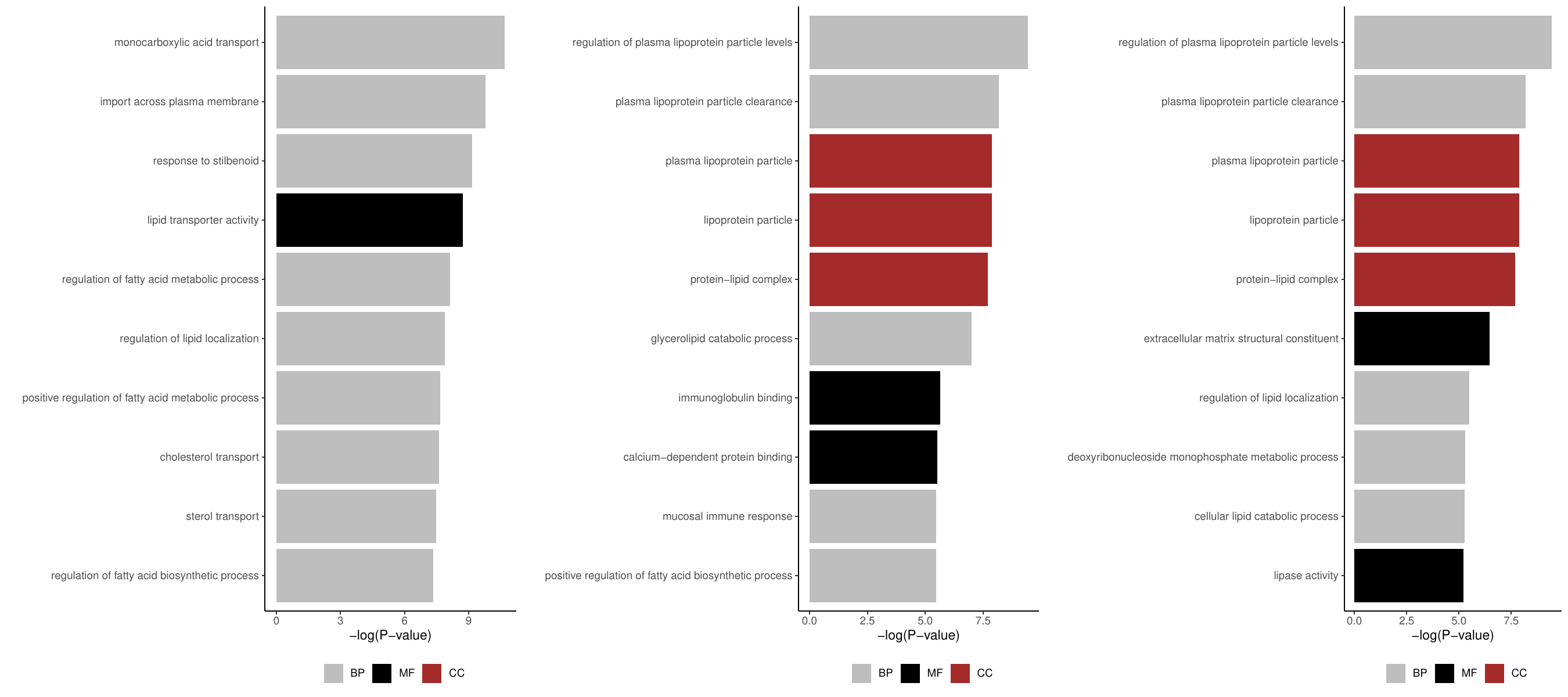}
\caption{Top ten GO terms (biological processes, cellular components and molecular functions)  obtained from the enrichment analysis for the genes selected by lasso\SIR (left), two-stage lasso (middle), and two-stage lasso\SIR (right).}
\label{fig:GO}
\end{figure}

In this subsection, we further interpret the biological consequences of the genes selected by different estimators as outlined in Section 7.1 of the main paper. To that end, we conducted a gene ontology enrichment analysis for the genes chosen by each method. By annotating these genes for their participation in gene ontology, this analysis aims to identify which biological functions are over-represented in the selected genes \citep{bleazard2015bias}. This over-representation is measured by the $p$-values, adjusted for multiple comparisons, of the null hypothesis that this gene set is randomly associated with the GO. A smaller $p$-value thus indicates a more significant biological function. Furthermore, the gene ontology is divided into three main groups: biological processes (BP), molecular function (MF), and cellular component (CC).          
Figure \ref{fig:GO} indicates that all the three methods capture a strong association between the obesity and lipid-related processes, including lipid transport, metabolism, and regulation. The one-stage lasso \SIR estimator focuses on genes regulating lipid transport and metabolic processes (e.g., lipid transporter activity, monocarboxylic acid transport).
 
By contrast, both the two stage lasso and two-stage lasso \SIR estimators emphasize plasma lipoprotein metabolism (e.g., plasma lipoprotein particle clearance, protein-lipid complex), with the former putting more emphasis on the interplay between lipid metabolism and immune regulation (e.g., immunoglobulin binding) while the latter focuses more on lipid catabolism and extracellular lipid transport (e.g., extracellular matrix structural constituent, lipase activity).   

\FloatBarrier

\subsection{NHANES data}
\begin{figure}[htb]
\centering
\includegraphics[width = 0.5\textwidth]{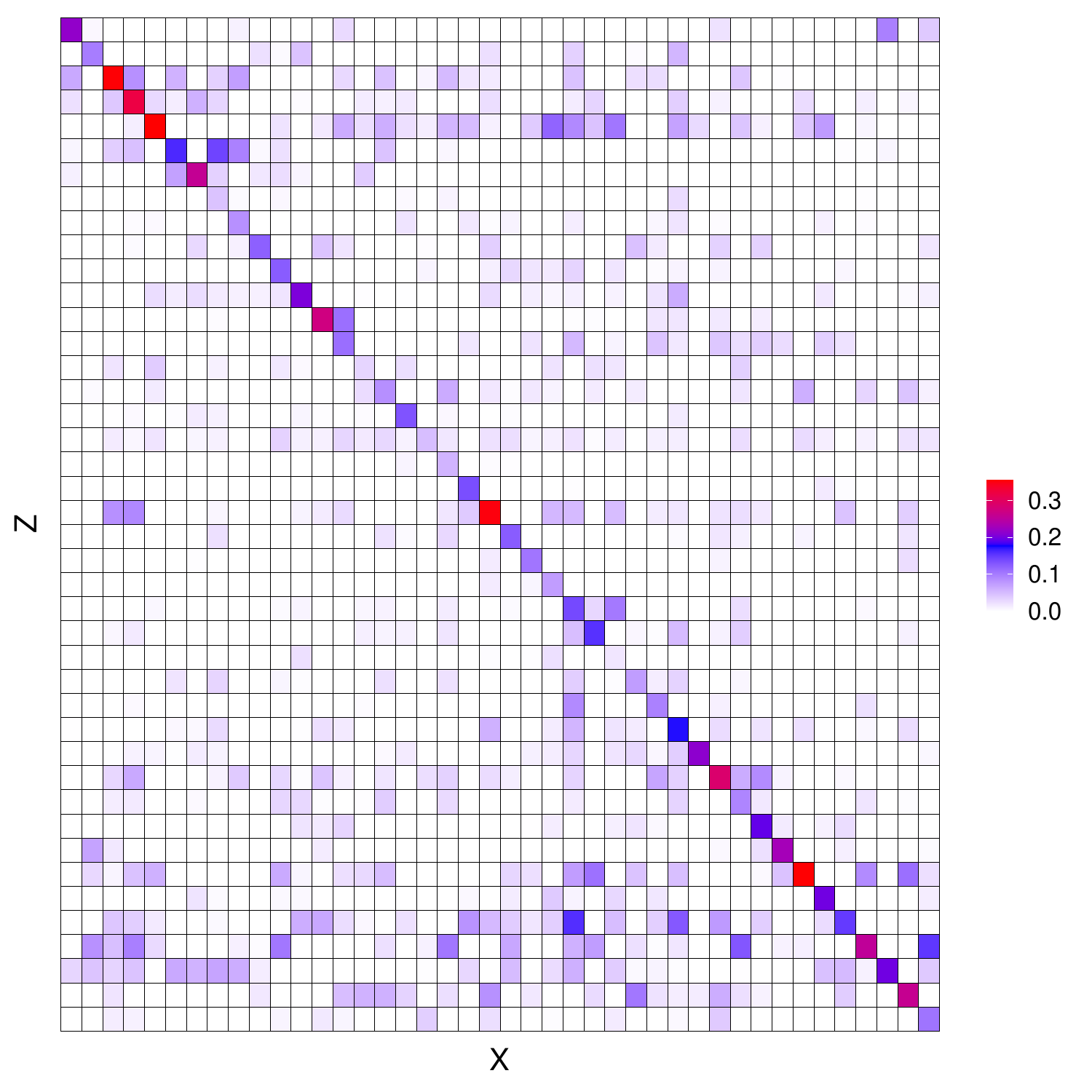}
\caption{Heatmap corresponding to the absolute value of the elements of $\widehat{\bGamma}$ from the first stage of the 2SLSIR estimator applied to the NHANES data. Each column represents the consumption of one dietary component in the first interview, which is treated as a covariate. Each column represents the consumption of one dietary component in the second interview, which is treated as an instrument.}
\label{fig:data2-firststage}
\end{figure}

\FloatBarrier

\end{document}